\documentclass[twocolumn]{aastex701}
\usepackage{apjfonts}
\usepackage{rotating}
\usepackage{amsmath, color,graphicx,tablefootnote,subfigure}
\usepackage{graphics,subfigure,hyperref,float}
\usepackage[rightcaption]{sidecap}

\newcommand\hi{\ion{H}{1}}
\newcommand\hii{\ion{H}{2}}
\newcommand\hei{\ion{He}{1}}
\newcommand\heii{\ion{He}{2}}
\newcommand\nii{[\ion{N}{2}]}
\newcommand\oi{[\ion{O}{1}]}
\newcommand\oii{[\ion{O}{2}]}
\newcommand\oiii{[\ion{O}{3}]}
\newcommand\neiii{[\ion{Ne}{3}]}
\newcommand\sii{[\ion{S}{2}]}
\newcommand\siii{[\ion{S}{3}]}
\newcommand\cliii{[\ion{Cl}{3}]}

\newcommand\ariii{[\ion{Ar}{3}]}
\newcommand\ariv{[\ion{Ar}{4}]}

\newcommand\feii{[\ion{Fe}{2}]}
\newcommand\feiii{[\ion{Fe}{3}]}

\newcommand\fev{[\ion{Fe}{5}]}
\newcommand\te{$T_e$}
\newcommand\den{$n_e$}
\newcommand\yp{$Y_{\rm p}$}

\newcommand{\multrow}[1]{\begin{tabular}{@{}c@{}} #1 \end{tabular}}

\shorttitle{The LBT $Y_{\rm p}$ Project II}
\shortauthors{Rogers et al.}
\accepted{to ApJ on June 4th, 2026}

\begin{document}

\title{The LBT $Y_{\rm p}$ Project II: MODS Spectra, Physical Conditions, and Oxygen Abundances in Local Metal-Poor Nebulae}

%
%

\author[0000-0002-0361-8223]{Noah S.\ J.\ Rogers}
\affiliation{Center for Interdisciplinary Exploration and Research in Astrophysics (CIERA), Northwestern University, 1800 Sherman Avenue, Evanston, IL, 60201, USA}
\email{noah.rogers@northwestern.edu}
\author[0000-0003-0605-8732]{Evan D.\ Skillman}
\affiliation{Minnesota Institute for Astrophysics, University of Minnesota, 116 Church St. SE, Minneapolis, MN 55455}
\email{skill001@umn.edu}
\author[0000-0003-1435-3053]{Richard W.\ Pogge}
\affiliation{Department of Astronomy, The Ohio State University, 140 W 18th Ave., Columbus, OH, 43210}
\affiliation{Center for Cosmology \& AstroParticle Physics, The Ohio State University, 191 West Woodruff Avenue, Columbus, OH 43210}
\email{pogge.1@osu.edu}
\author[0009-0006-2077-2552]{Erik Aver}
\affiliation{Department of Physics, Gonzaga University, 502 E Boone Ave., Spokane, WA, 99258}
\email{aver@gonzaga.edu}
\author[0000-0003-4912-5157]{Miqaela K.\ Weller}
\affiliation{Department of Astronomy, The Ohio State University, 140 W 18th Ave., Columbus, OH, 43210}
\email{weller.133@buckeyemail.osu.edu}
\author[0000-0002-4153-053X]{Danielle A. Berg}
\affiliation{Department of Astronomy, The University of Texas at Austin, 2515 Speedway, Stop C1400, Austin, TX 78712, USA}
\email{daberg@austin.utexas.edu}
\author[0000-0001-8483-603X]{John J. Salzer}
\affiliation{Department of Astronomy, Indiana University, 727 East Third Street, Bloomington, IN 47405, USA}
\email{josalzer@iu.edu}
\author[0000-0002-2901-5260]{John H. Miller, Jr.}
\affiliation{Minnesota Institute for Astrophysics, University of Minnesota, 116 Church St. SE, Minneapolis, MN 55455}
\email{mill9614@umn.edu}
\author[0009-0009-2024-9317]{Jayde Speigel}
\affiliation{Department of Astronomy, The Ohio State University, 140 W 18th Ave., Columbus, OH, 43210}
\email{spiegel.85@buckeyemail.osu.edu}
\author[0000-0001-6369-1636]{Allison L. Strom}
\affiliation{Center for Interdisciplinary Exploration and Research in Astrophysics (CIERA), Northwestern University, 1800 Sherman Avenue, Evanston, IL, 60201, USA}
\affiliation{Department of Physics and Astronomy, Northwestern University, 2145 Sheridan Road, Evanston, IL, 60208, USA}
\email{allison.strom@northwestern.edu}


\begin{abstract}

Empirically measuring the primordial He mass fraction, \yp, requires a significant number of low-metallicity nebulae with direct constraints on He/H and O/H abundances. This technique requires high-fidelity measurements of the gas-phase physical conditions, namely the electron temperature (\te) and density (\den). To this end, we present deep rest-optical spectroscopy for a sample of 62 low-metallicity ($\lesssim$ 20\% solar O/H) galaxies acquired using the Multi-Object Double Spectrographs (MODS) on the Large Binocular Telescope (LBT) as part of the LBT \yp\ Project. We discuss new fitting methods that recover the intensity of up to 61 H and He recombination lines, of which, up to 26 will be used to determine gas-phase He abundances, and we examine the emission line properties of the LBT \yp\ Project sample. We assess different scaling relations in the low-metallicity interstellar medium (ISM), finding that \den\ariv\ measured in 31 targets is systematically larger than \den\sii\ or \den\oii.
The larger densities are insufficient to significantly bias \te\oiii\ or the O/H abundance. \te\siii\ and \te\oiii\ are strongly correlated over a range of $\sim$10$^4$ K with very low scatter, and we calibrate new \te\siii-\te\oiii\ scaling relations for use in other low-metallicity environments. We examine different \te\ measured in the low-ionization gas, finding significant scatter compared to \te\oiii.
The precision direct O/H derived in this analysis (median uncertainty $\sim$4\%) are consistent with prior literature measurements, albeit with relatively large scatter. These data provide a key component necessary to empirically measure \yp\ and the abundance patterns of other elements in the ISM.

\end{abstract}


\section{Introduction}\label{sec1}

The ideal measurement of the Primordial Helium (\yp) abundance would come from a measurement of He/H from the early universe, but the present CMB-only constraints on \yp\ are relatively weak \citep[e.g.,][]{Planck2020}.  Thus, the current favored methodology follows that of \citet{Peimbert1974, Peimbert1976}, where the helium abundance is measured as a function of oxygen abundance and extrapolated back to an oxygen abundance of zero.
It follows that robust measurements of the oxygen abundance (or some other suitable heavy element) are required in addition to the helium abundance for the determination of \yp. 

High-fidelity measurements of the O abundance are possible in the ionized phase: the rest-frame optical spectrum of \hii\ regions and star-forming galaxies contains many strong collisionally-excited lines (CELs) from metal ions and recombination lines of H and He. From these optical spectra, it is possible to measure the physical conditions of the electron gas, the ionic abundance of O$^+$ and O$^{2+}$, and numerous \hi\ and \hei\ lines necessary to constrain He/H \citep[e.g.,][]{aver2021}. The optical spectra that are acquired for this analysis have many other related uses. For example, measurements of the multi-phase electron temperature (\te) and density (\den) structure of the interstellar medium (ISM) are used for comparing to the temperatures determined solely from H and He emission lines. Other elements (e.g., N, Ne, S, Ar) available in the optical assess systematic uncertainties related to the use of oxygen as the baseline element for comparing He abundances \citep{fern2018}.
Finally, optical CELs are useful for assessing the ionizing sources within a galaxy \citep[e.g.,][]{bald1981}, which can be used to exclude nebula with significant Active Galactic Nuclei (AGN) or shock ionization where the techniques used to measure He/H are inappropriate to apply.

Therefore, a precision measurement of \yp\ requires a statistically significant sample of low-metallicity galaxies with high-quality rest-frame optical and near-infrared (NIR) spectroscopy capable of measuring the physical conditions and chemical abundances in the ISM. In addition, because of the requirement of relatively high spectral resolution and a large wavelength baseline, these optical spectra are sensitive to a wealth of relatively weak lines that are not regularly detected in typical spectra collected for the purpose of nebular abundances. In this regard, these spectra share a commonality with the spectra used to obtain abundances from the very faint optical recombination lines \citep[e.g.,][]{esteban2005,este2009,esteban2014,este2020,skil2020}. 

To this end, the Large Binocular Telescope (LBT) Primordial He Abundance Project, hereafter LBT \yp\ Project, 
has conducted a dedicated observing campaign to acquire optical and NIR spectroscopy of over 60 local star-forming galaxies and metal-poor \hii\ regions. As described in \citet[][hereafter Paper I]{Skillman2026}, the targets are well-known systems such as the extremely metal-poor galaxies Leo P \citep{giov2013,skil2013}, I Zw 18 \citep{sear1972,skil1993}, and DDO 68 \citep{anni2019}, to more extreme emission line systems like Mrk 71 \citep{gonz1994,mich2017,chen2023} and SBS0335$-$52E \citep{izot1990,woff2021}. While He/H abundances have been measured in some of these nebulae, the inhomogeneous nature of the archival observations (i.e., telescope/instrument configuration, wavelength coverage, and available \hi\ and \hei\ lines) results in significant systematic uncertainties when attempting to measure $Y$ vs.\ O/H.

The LBT \yp\ Project's Multi-Object Double Spectrograph \citep[MODS;][]{pogg2010} data provide the wavelength coverage and requisite spectral resolution to measure a plethora of \te\ and \den\ diagnostics. As has been demonstrated in other metal-rich nebulae \citep{berg2015,crox2016} and metal-poor galaxies \citep{skil2013,anni2019,berg2021}, MODS spectra are capable of measuring \te\ and \den\ from ions spanning 10.4 -- 40.7 eV in ionization potential (IP), which are required for determining the ``direct'' abundance \citep{dine1990,maio2019} of O in the multi-phase ISM. These direct abundances are relatively free of systematic uncertainties that characterize ``strong-line'' abundances \citep[see][]{kewl2008,mous2010} and provide a necessary component for determining the relation between He/H and gas-phase metallicity. The utility of MODS for the measurement of He abundance has been illustrated in our pilot analyses of the extremely metal-poor galaxies Leo P \citep{aver2021} and AGC198691 \citep{aver2022}. We now provide a homogeneous sample of low-metallicity galaxies with MODS spectroscopy, enabling the robust determination of \yp.

A primary motivation of the LBT \yp\ Project is to obtain spectra of the lowest metallicity objects to lessen the dependence on the assumption of a linear relationship between $Y$ and O/H.  In short, we would like the observations at the lowest metallicity to have the highest impact on the determination of \yp, not the more metal-rich objects that play a dominant role in determining the slope of the $Y$ vs.\ O/H relationship. Thus, as a byproduct of our new determination  of \yp, we can examine scaling relations in the metal-poor ISM, particularly between gas-phase physical conditions like \te\ and \den. While \te\ scaling relations have been calibrated on the temperature trends of local star-forming regions \citep{este2009,pily2009,crox2016,arel2020,berg2020,roge2021,rick2024}, the shape of these different relations can diverge both from other empirical calibrations and from photoionization model predictions \citep[e.g.,][]{camp1986,garn1992,lope2012}.
Understanding the evolution of \te\ in the ISM at low-metallicity is crucial for reliable abundance surveys in the local Universe and for interpreting the emerging direct \te\ in high-\emph{z} galaxies observed with JWST \citep{roge2024,roge2026cecilia,cataldi2025,sanders2025,welc2025}. Many of these high-\emph{z} galaxies also exhibit elevated \den\ in the low-ionization gas \citep{sand2016,stro2017,topping2025}, and recent observations of UV emission lines in high-\emph{z} extreme emission line galaxies (EELGs) indicate the presence of very high-density gas \citep[$>$ 10$^6$ cm$^{-3}$, see][]{topp2024}. Deep optical spectra of low-metallicity, highly-ionized galaxies at \emph{z}$=$0 assess both the density stratification in different ionization zones of the ISM and hidden high-density condensations that may bias \te\ and abundance measurements \citep{mend2023}, which may be important considerations in galaxies at high \emph{z}.

The format of this paper is organized as follows. In \S\ref{s:data} we describe the MODS data reduction and analysis, ultimately delivering the \hi\ and \hei\ emission line fluxes and equivalent widths (EWs) for the He abundance analysis. In \S\ref{s:phys}, we measure the gas-phase physical conditions, and we compare \den\ measured from different ions in the ISM. We assess the \te\ trends in the LBT \yp\ Project nebulae in \S\ref{s:tescaling}, providing new \te\ scaling relations between \te\oiii\ and \te\siii. In \S\ref{s:loh}, we measure the total gas-phase O/H in the LBT \yp\ Project sample and compare these to other direct measurements available in the literature. We summarize our conclusions in \S\ref{s:conclusions}.

\section{Data Reduction and Analysis}\label{s:data}

\subsection{Observations and Reduction}

Paper I details the target selection criteria, observation setup, and aspects of the MODS data reduction up to the 1D spectra of each metal-poor nebula. While we refer the reader to this paper for these specifics, we briefly highlight the standard reduction procedure here. All science targets are observed with a 1\farcs0$\times$60\farcs0 long slit. MODS1 and MODS2 have a spectral resolution of R$\sim$1850 and 2300 in the blue and red channel, respectively, and a combined wavelength coverage of 3300 -- 10000 \AA. For all data acquired within a given observing run, we use the \textsc{modsCCDRed} \citep{pogg2019} programs to process (bias subtract and flat field correct) the raw CCD images of the science targets, standard stars (at least one observed per night), and calibration lamps (observed once per run). Cosmic rays are removed from the science images using the IDL implementation of \textsc{lacosmic} \citep{vanD2001}, which are then average combined. Standard star images are median combined, where no difference in the flux calibration was observed when using an average combination of the standard star observations. The choice of standard star is made based on the time of observation in relation to the science targets and the conditions during observation.

The \textsc{modsIDL} \citep{crox2019} reduction package is used to reduce the science, standard, and lamp images, producing wavelength solutions for the 2D spectra. Standard stars are extracted and used to determine the shape of the sensitivity function for the MODS1 and MODS2 detectors. Local sky subtraction is performed using the available sky in the longslit spectrum, where the sky is modeled with a B-spline function and subtracted from the 2D science image. Science targets are extracted using a range of widths based on the slit profiles of the 2D spectra centered on H$\beta$ and H$\alpha$ for the blue and red channel, respectively. We extract every object using at least two extraction widths and check that all line ratios and physical conditions are consistent within uncertainty; when the extractions are consistent within uncertainty, we adopt the narrow extraction to reduce the RMS noise in the continuum. When wide and narrow extractions disagree, we select the wider extraction to ensure that no significant emission is omitted, except in instances where a neighboring object is affecting the flux reported for the primary target.

\begin{figure*}[!t]
\epsscale{1.0}
   \centering
   \includegraphics[width=0.94\textwidth]{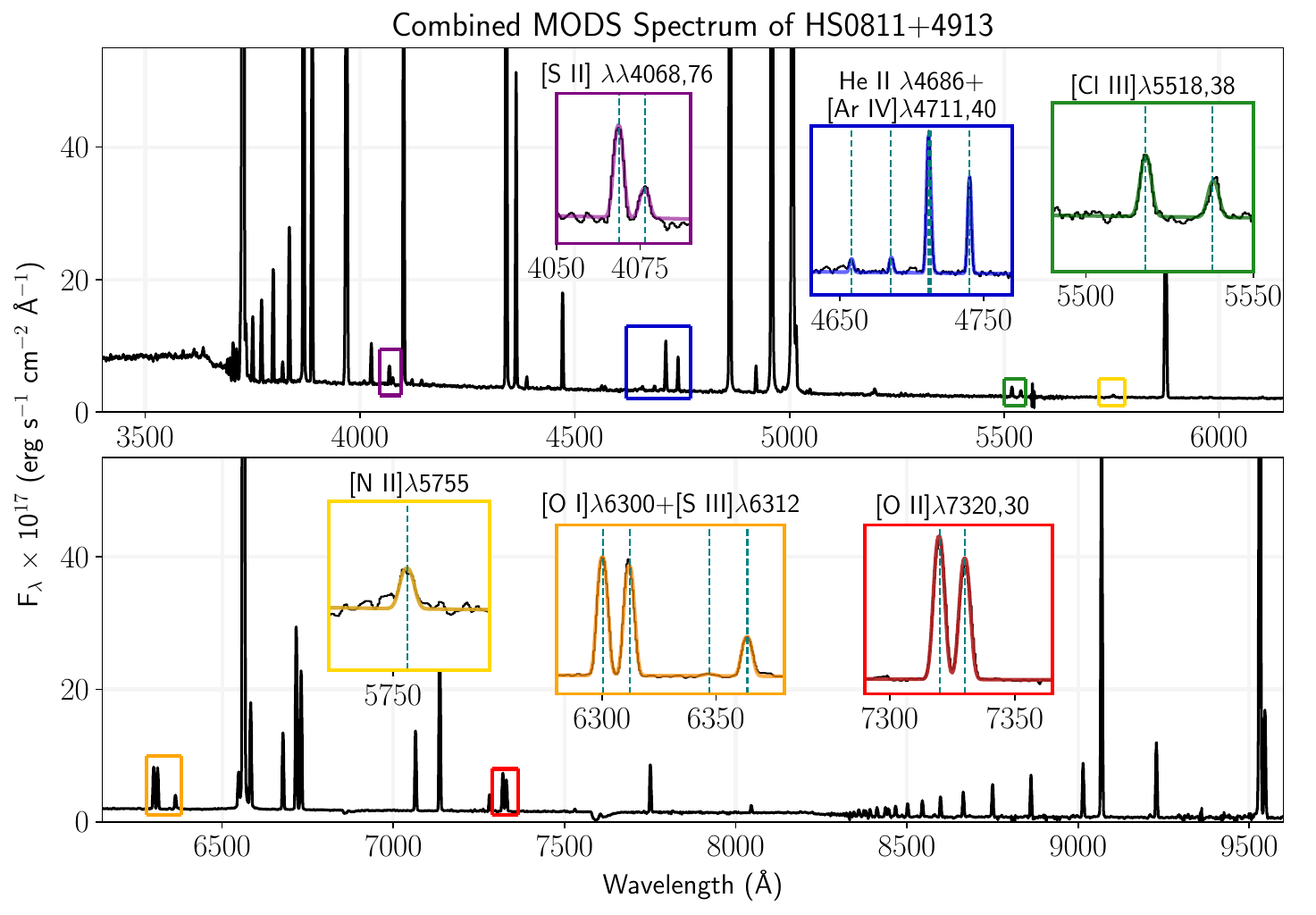}
   \caption{Combined MODS spectrum of the low-metallicity galaxy HS0811+4913. Faint \te- and \den-sensitive emission lines are highlighted in the subplots. The \te-sensitive auroral line of \oiii$\lambda$4363 is clearly visible without magnification. The spectrum shows a distinct Balmer and Paschen jump, along with numerous intense and faint \hei\ lines crucial for the determination of He$^+$/H$^+$.}
   \label{fig:spectrum}
\end{figure*} 

With the individual spectra from the MODS1 and MODS2 channels, we combine the blue and red spectra at rest-frame 5650 \AA\ using the \textsc{spectres} package \citep{carn2017} to preserve the native sampling of MODS (0.5 \AA\ per pixel). We also examine the agreement of the blue and red spectra around the dichroic crossover and fit a linear function to the continuum before and after 5650 \AA. In instances where there is an offset in the continua (a trend that can be produced by small differences in sky subtraction between the blue and red spectra), we apply an additive constant to the red spectrum to produce an agreement at the dichroic (9 objects in total). There are instances where the MODS2 red channel spectra are a constant factor greater than the MODS1 red channel spectra and do not align with the MODS2 blue spectra at the dichroic (8 objects in total). For these objects, we apply a multiplicative factor to the MODS2 red spectra to produce agreement at the dichroic. When MODS1 and MODS2 data are available for a given object, we combine the two using a variance-weighted average. To illustrate the quality of the MODS sample, we select a galaxy with a F(H$\beta$) below the sample median and plot the combined MODS spectrum in Figure \ref{fig:spectrum}. The MODS spectrum of HS0811+4913 reveals the high signal-to-noise (S/N) detections of many faint emission lines crucial for \te, \den, and ionic abundance measurements.


The final merged 1D MODS optical spectra (wavelengths, relative fluxes, and flux errors) will be available on Zenodo (\!\dataset[DOI: 10.5281/zenodo.19699768]{https://doi.org/10.5281/zenodo.19699768}; \citealt{mods1d:zenodo} in FITS binary table format.

\subsection{Emission Line Fitting}\label{s:linefit}

Both the present chemical abundance analysis and the determination of He/H \citep[][hereafter Paper IV]{Aver2026} require reliable measurements of emission lines spanning the full optical spectrum. To fit the emission lines, we split the spectrum into 18 windows and assume that the stellar continuum in each window can be approximated with a linear continuum over the small wavelength range. We do not consider areas of the continuum with significant stellar absorption in the linear continuum fit. The contribution of stellar absorption to the total Balmer line flux is accounted for via the iterative reddening correction (see \S\ref{s:redcorr}).

\begin{figure*}[!t]
   \centering
   \includegraphics[width=0.50\textwidth]{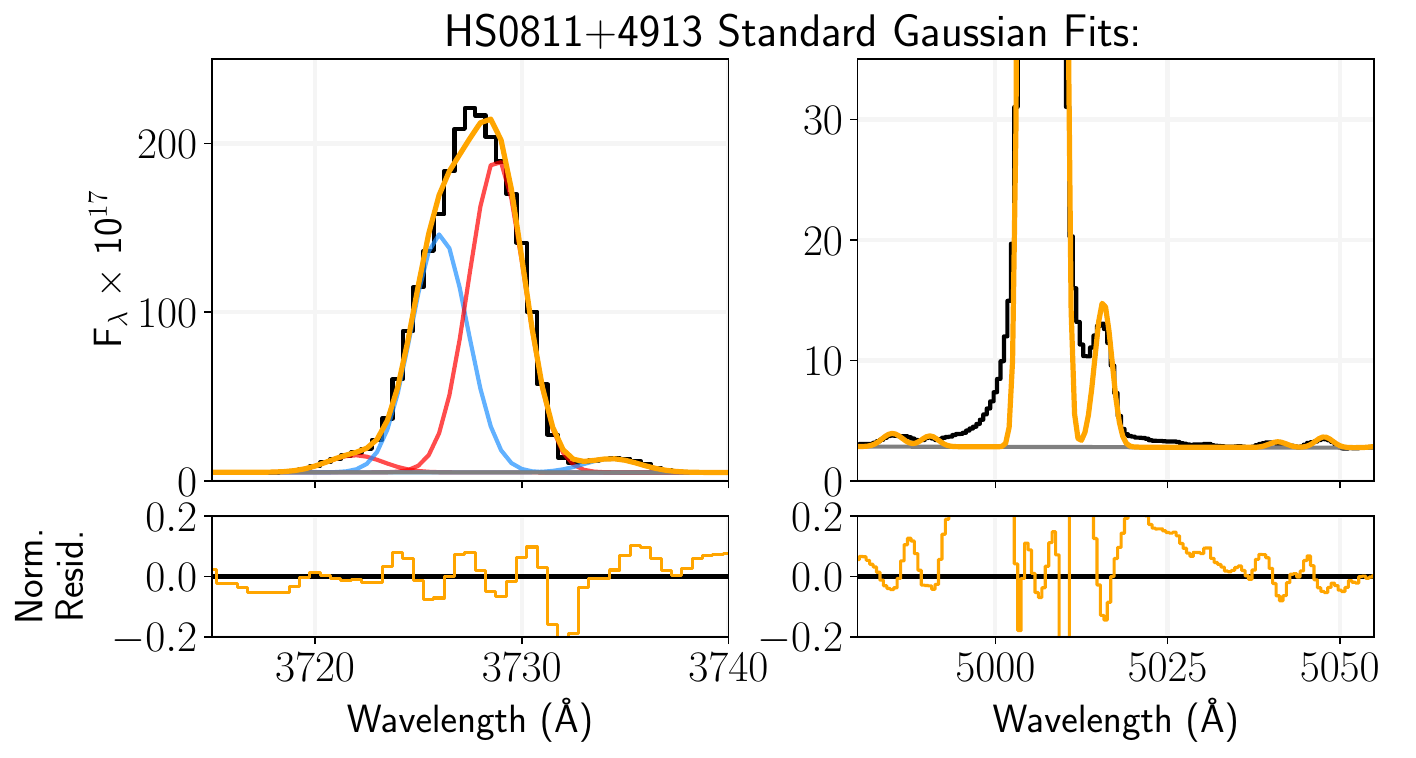}
   \hskip -1ex
   \includegraphics[width=0.50\textwidth]{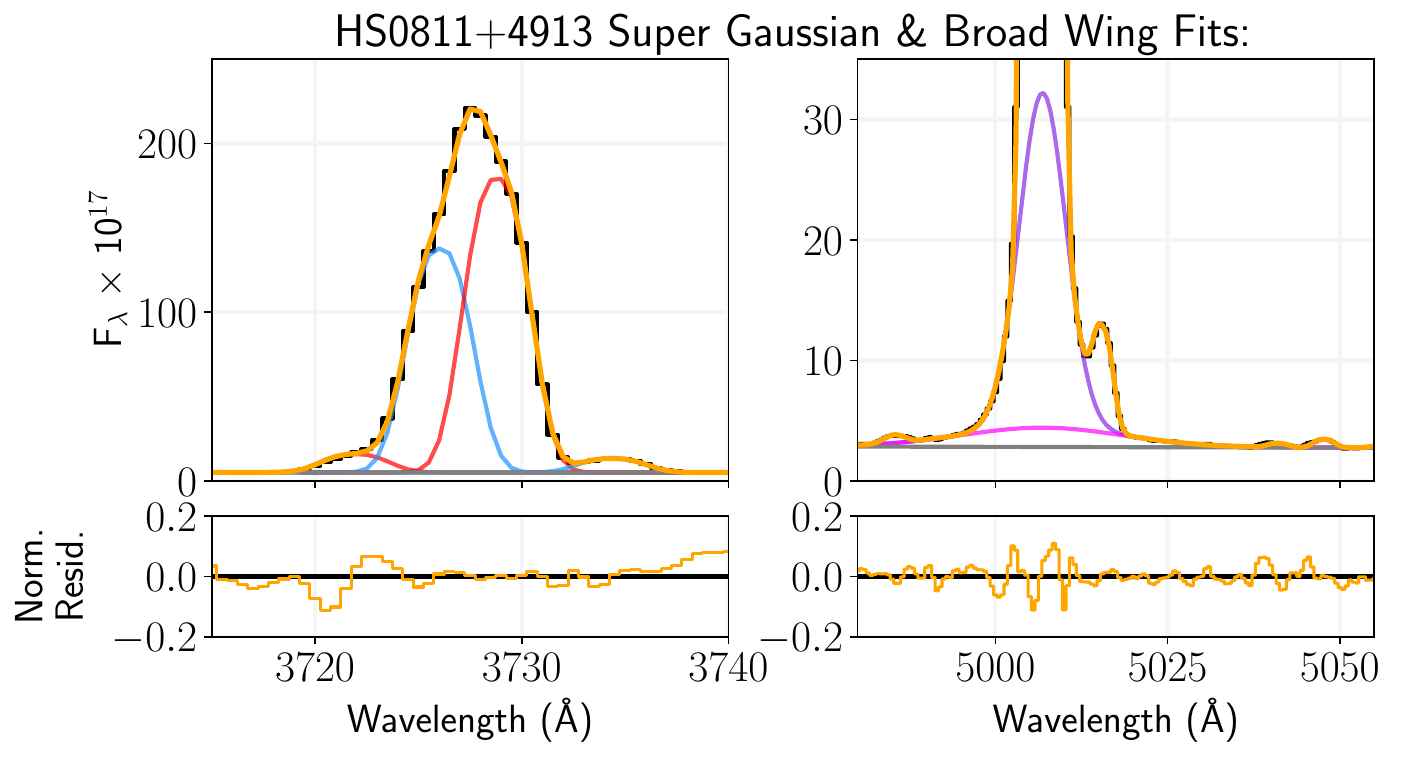}
   \caption{Examples of emission line fits in the galaxy HS0811+4913. The left two panels uses a combination of a linear continuum plus Gaussian emission line profiles to fit the H14, H13, and \oii$\lambda\lambda$3726,29 (first panel) and \oiii$\lambda$5007 and \hei$\lambda$5016 (second panel). The 1D spectrum is plotted in black, the full fit is provided in orange, individual narrow emission lines are in alternating blue and red. The standard Gaussian fits have normalized residuals (lower panels) that can exceed 20\%. The right two panels plot new fits using super-Gaussian profiles for the emission lines. This provides significantly better fits to the blend of \hi\ and \oii\ lines in the third panel. The inclusion of a two-component broad wing on \oiii$\lambda$5007 (fourth panel, shown in pink and purple) provides a reliable fit to both the strong \oiii\ line and the faint \hei\ line.
   }
   \label{fig:sg_comp}
\end{figure*}

Emission lines are generally approximated with Gaussian functions based on the combination of thermal, Doppler, and instrumental broadening of an emission line \citep[e.g.,][]{fern2024}. 
However, the emission line profiles observed in many of the MODS1 and MODS2 spectra are not well characterized by a Gaussian profile. This is primarily due to the targets being extended relative to the slit width and thus the instrumental profile reflects the boxcar-like nature of the slit.  This presents a challenge for simultaneously fitting faint emission lines adjacent to intense lines, as exhibited in the leftmost panel of Figure \ref{fig:sg_comp}. For example, the shape of the blended \oii\ lines in HS0811+4913 is not accurately reproduced by the Gaussian emission line fits, leading to normalized residuals that can exceed 20\% of the measured flux. In this case, the residuals are sufficiently large from the mismatch of the observed and fit line profiles that the weak H13 recombination line is poorly recovered.

To fit the emission lines, we adopt a super-Gaussian emission line model with functional form:
\begin{equation}
    F_\lambda = \mbox{A} \times e^{-\frac{1}{2}(\frac{\left|{\lambda - \lambda_0}\right|}{\sigma_{SG}})^P},
    \label{eq:sg}
\end{equation}
where A is the amplitude of the super-Gaussian, $\lambda_0$ is the central wavelength of the emission line, and $P$ is the super-Gaussian power. $\sigma_{SG}$ is the width parameter of the super-Gaussian,
\begin{equation}
    \sigma_{SG} = \frac{\mbox{FWHM}}{2\times(2\times\mbox{ln}(2))^{1/P}},
    \label{eq:sg_sig}
\end{equation}
where FWHM is the Full-Width at Half Maximum of the super-Gaussian \citep{beir2017}. Assuming the emission lines can be described by Equation \ref{eq:sg} allows for a flexible fit to the shape of the line profile, particularly the peak of the emission line. Equation \ref{eq:sg} reduces to a Gaussian profile when $P=2$, but $P<2$ produces a line with a sharper peak and $P>2$ results in a flat-topped line profile. At the extremes, the line profile becomes a discontinuous peak with broad wings as $P\rightarrow1$ and a top hat profile at $P\gg2$.

Within each fitting window, we adopt the linear continuum model and assume all narrow lines have constant super-Gaussian FWHM, $P$, and wavelength offset from the theoretical wavelength, but we fit for each unique super-Gaussian amplitude. The number of free parameters, therefore, is $N+3$, where $N$ is the number of emission lines fit within each window. We then construct a $\chi^2$ function to minimize using \textsc{scipy} \citep{2020SciPy} and determine the best-fit solution for all emission lines within the window. To calculate the emission line flux, we integrate the best-fit super-Gaussian profile over the wavelength grid of the full MODS spectrum. The super-Gaussian fit to the \oii\ lines of HS0811+4913 is provided in the third panel of Figure \ref{fig:sg_comp}. The new approach reduces the normalized residuals below 10\% across the full line profile and reproduces the \oii\ doublet, allowing for a reliable estimation of the electron density in the low-ionization gas. Further, faint lines such as H13 and H14 are well fit with the same super-Gaussian FWHM and $P$ parameter. Note that the remaining residual at $\sim$3722 \AA\ is due to the blend of H14 with the faint \siii$\lambda$3722 line, the latter of which is not independently fit.

A subsample of the metal-poor nebulae shows broad wings on the intense H$\beta$, H$\alpha$, and \oiii\ optical lines. These features can further complicate the Gaussian fitting procedures, as illustrated in the second panel of Figure \ref{fig:sg_comp}. In this example, the \hei\ $\lambda$5016 line flux is greatly overestimated and the normalized residuals around \oiii$\lambda$5007 are well above 20\%. This is also true for the flux of the \nii\ strong lines, which are faint in metal-poor galaxies and, therefore, are sensitive to how the broad component of H$\alpha$ is treated. The broad components may be related to high-velocity gas in the ISM, or are due to differences in the instrumental line-spread function from the assumed super-Gaussian line profile. The latter is consistent with the appearance of the broad components on the brightest emission lines.
Prior LBT/MODS studies have reported the presence of such broad features in similarly metal-poor objects observed with MODS \citep{izot2017,berg2021}, and have found that the broad component generally accounts for a small fraction of the overall line flux. Given the potential diagnostic importance of \hei\ $\lambda$5016 and the \nii\ strong lines, we model the broad wings to obtain more robust fits to both the strong lines and neighboring faint lines.

When there is clear evidence of broad emission, we follow the methodology of \citet{berg2021} and fit the broad wing with a two-component Gaussian model. The slightly-broad Gaussian component captures slight deviations of the line profile from the narrow super-Gaussian, while the very-broad Gaussian models both deviations from the super-Gaussian profile and the contribution of high-velocity gas ($v>1000$ km/s) that can affect the weak lines. The flux of the slightly-broad component accounts for 1-3\% of the narrow component, and the narrow and slightly-broad components are added together when reporting the total line flux. The very-broad component of \oiii$\lambda$5007 is always $<$ 1.5\% of the flux of the narrow super-Gaussian, while the broad component of H$\alpha$ is typically $<$ 2\% of F(H$\alpha$) for the full sample. As this component may be associated with a high-velocity outflow that does not have the same physical conditions as the \hii\ region, we exclude the flux of the very-broad component from the reported total line flux.

Finally, the measurements of $T_e$ and ionic abundance are sensitive to the flux of very faint emission lines, such as auroral \oiii$\lambda$4363 and \hei$\lambda$4026. To verify the super-Gaussian fit, we fit faint emission lines by hand using the \textsc{IRAF}\footnote{NOIRLab IRAF is distributed by the Community Science and Data Center at NSF NOIRLab, which is managed by the Association of Universities for Research in Astronomy (AURA) under a cooperative agreement with the U.S. National Science Foundation.} \textsc{splot} routine. When there is significant disagreement between the automatic and \textsc{splot} line fluxes, we adopt the \textsc{splot} fluxes for our analysis. Corrections are often needed for emission lines that are close to stellar absorption features and areas in the NIR with large sky subtraction residuals, both of which can bias the linear continuum fits.

We consider three sources of uncertainty when reporting the error in emission line flux. The first, $2\times RMS \times\sqrt{2\times FWHM}$, is related to the noise in the continuum around the emission line that could affect the super-Gaussian fit \citep[see][]{berg2013}. The second term is a constant 1\% uncertainty, which is the minimum uncertainty determined by the level of fixed pattern noise present in the MODS detectors. The third term is the flux calibration of the MODS detectors themselves, and we refer the reader to Paper I for details regarding these response function uncertainties. In brief, MODS observations of the standard stars G191B2b and GD153 spanning three years (2021-2024) are used to derive the stability of the MODS response functions. Blue and red channel response function uncertainties are derived as a function of wavelength for both MODS1 and MODS2. We sample these functions at the observed wavelength of a given emission line to retrieve the uncertainty in the MODS flux calibration at that wavelength. This percent uncertainty is added in quadrature with the first two terms to produce the final reported line flux uncertainty. We consider an emission line detected if its S/N is above 3.

\subsection{Reddening Correction}\label{s:redcorr}

After emission line fitting, line fluxes and EWs of 17 \hi\ and 9 He recombination lines are taken as inputs to the He abundance MCMC code (see Paper IV for more details). More \hi\ and \hei\ lines are available and fit in the MODS optical spectra (up to 31 \ion{H}{1} lines, 28 \hei\ lines, and 2 \heii\ lines), but are not detected in a sufficient portion of the sample. The addition of these other \hi\ and \hei\ lines may help provide better constraints on the resulting He$^+$/H$^+$ abundance, but are currently not included in the \yp\ analysis. The MCMC method in Paper IV solves for reddening independently, but for this analysis we use ratios of the four strongest Balmer emission lines in an iterative approach to infer $E(B-V)$ and $a_H$, the EW of the underlying stellar absorption in the Balmer lines. We solve for a single $a_H$ at H$\beta$ and scale the absorption EW for H$\delta$, H$\gamma$, and H$\alpha$ based on the ratio of Balmer EWs measured from the BPASS \citep{eldr2009,eldr2017,stan2018} spectral energy distributions for young stellar populations \citep[see][]{aver2021}.

In the iterative approach, as a first pass, the theoretical H$\alpha$/H$\beta$, H$\gamma$/H$\beta$, and H$\delta$/H$\beta$ ratios are calculated at \te$=$$10^{4}$ K and \den$=$$10^2$ cm$^{-3}$ using the atomic data of \citet{stor1995}. Next, the combination of $E(B-V)$ and $a_H$ that produces agreement between the observed and theoretical ratios are determined through a $\chi^2$ minimization using \textsc{scipy optimize}. We use the \citet{card1989} attenuation law for the calculation of $E(B-V)$. Holding $E(B-V)$ constant, $a_H$ is varied around the best-fit solution to determine the value at which the reduced $\chi^2$ value increases by 1; that value is taken as the uncertainty on the underlying absorption. A similar process is done for $E(B-V)$ by holding $a_H$ constant at the best-fit value.

\begin{figure*}[t]
   \centering
   \includegraphics[width=0.99\textwidth]{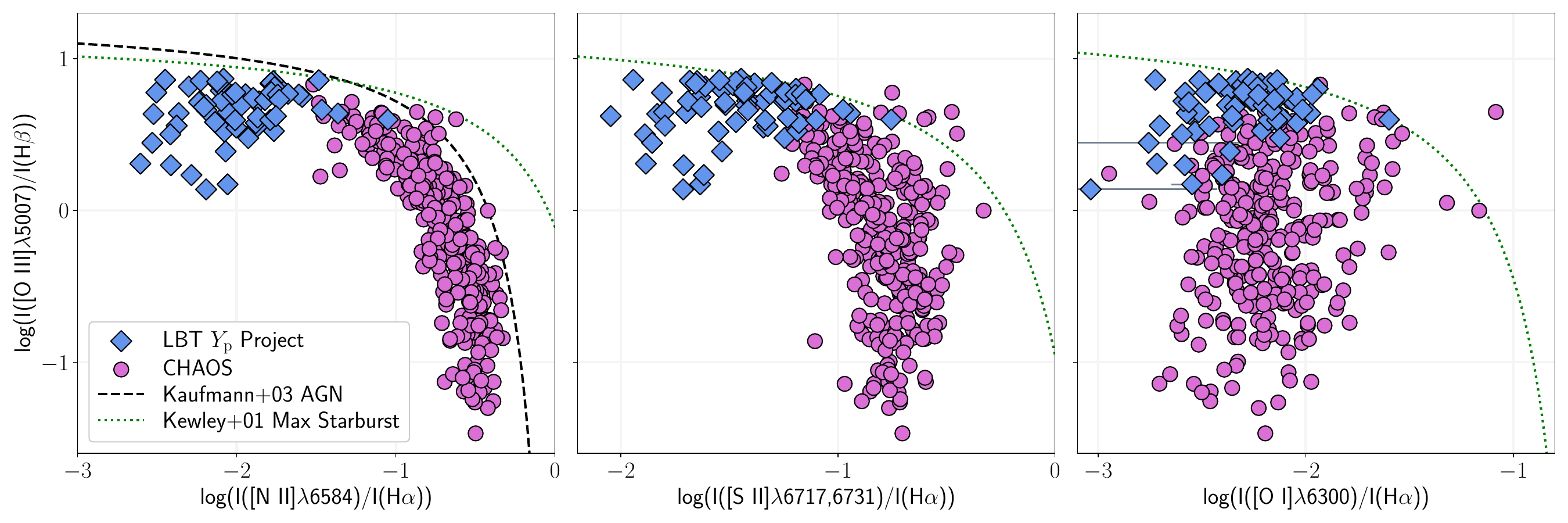}
   \caption{Emission line diagrams proposed by \citet{bald1981} and \citet{veil1987}: log(I(\oiii$\lambda$5007)/I(H$\beta$)) vs.\ log(I(\nii$\lambda$6584)/I(H$\alpha$)) (Left), log(I(\sii$\lambda$6717,31)/I(H$\alpha$)) (Center), and log(I(\oi$\lambda$6300)/I(H$\alpha$)) (Right). Line intensity ratios measured in the LBT \yp\ Project sample are plotted as blue diamonds. Line ratios measured in relatively metal-rich \hii\ regions from the CHAOS project are plotted as purple circles for reference. The \citet{kewl2001} and \citet{kauf2003} demarcations proposed to distinguish between stellar and AGN ionizing sources are plotted as black dashed and green dotted lines, respectively. The LBT \yp\ Project targets are characterized by highly-ionized, low-metallicity ISM and have line ratios consistent with extreme stellar ionization.}
   \label{fig:bptfigs}
\end{figure*}
\begin{figure*}[t]
   \centering
   \includegraphics[width=0.99\textwidth]{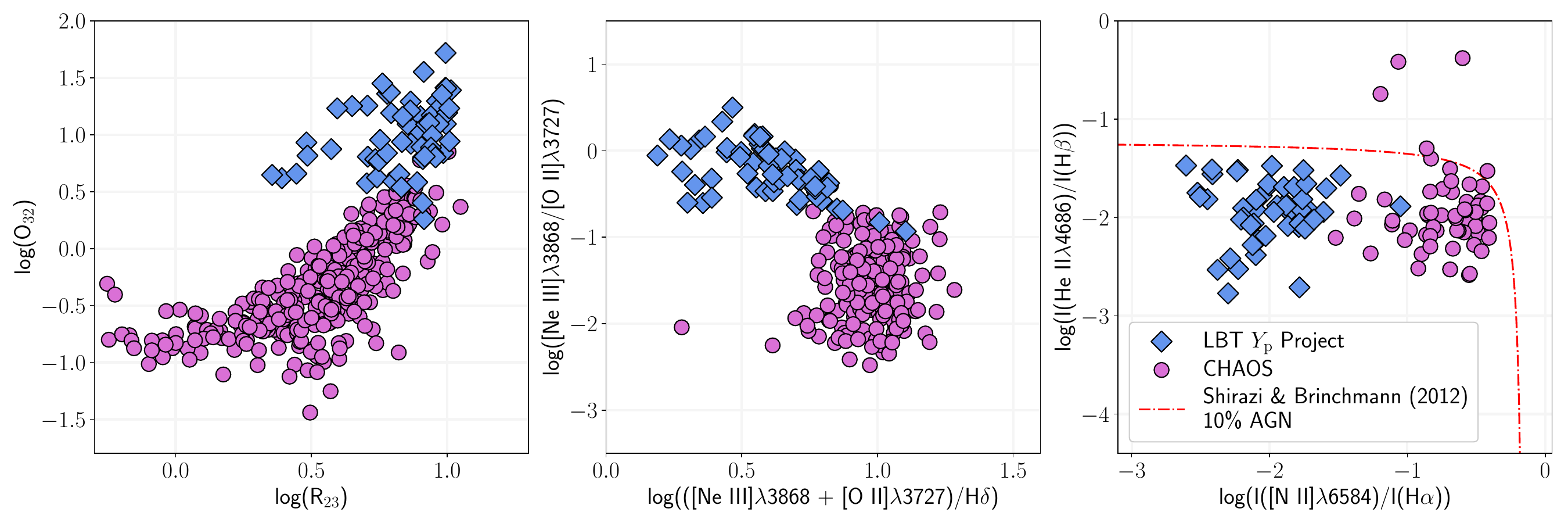}
   \caption{Additional line ratio diagrams to highlight the highly-ionized ISM of the LBT \yp\ Project targets. The shape and color of each point is consistent with the labeling from Figure \ref{fig:bptfigs}. Left: log(O$_{32}$) vs.\ log(R$_{23}$), the ionization-excitation diagram. Center: log(I(\neiii$\lambda$3869)/I(\oii$\lambda$3727)) vs.\ log(I(\neiii) + I(\oii)/I(H$\delta$)), an alternative ionization-excitation diagram. Right: log(I(\heii$\lambda$4686)/I(H$\beta$)) vs.\ log(I(\nii$\lambda$6584)/I(H$\alpha$)), a diagram to highlight nebular \heii\ emission. The red dot-dashed line indicates line ratios from 10\% AGN ionization \citep[from][]{shir2012}. All diagrams highlight the high ionization and excitation conditions observed in the targeted sample, while the line ratios remain consistent with predominantly stellar ionization.}
   \label{fig:extra_bptfigs}
\end{figure*}

With these two parameters, the spectrum is corrected for reddening, and underlying absorption is accounted for in the four strongest Balmer lines. \te\oiii\ is calculated using the intensity ratio of \oiii$\lambda$4363/\oiii$\lambda$5007, and this \te\ is used to update the theoretical ratios of the \hi\ lines. We iterate this procedure, calculating the new best-fit values for $E(B-V)$ and $a_H$ and their associated uncertainties at each step, until the change in the measured \te\oiii\ is $<$ 20 K. At the point of convergence, we use the final $E(B-V)$ to reddening correct the spectrum. We also account for $a_H$ in the four strongest Balmer lines, and we update the lower bounds on the uncertainties for $a_H$ such that the minimum $a_H$ cannot go below 0 \AA.
The observed line fluxes and EWs required to measure \yp\ are provided in Table \ref{t:hiinten}, as well as the best-fit $E(B-V)$ and $a_H$.
The uncertainty in a line intensity is determined using the line flux and reddening correction uncertainties. When reporting uncertainties on line intensity ratios, we consider the uncertainty in individual line fluxes, the error on $E(B-V)$, and the wavelength separation between the lines of interest. This approach is also adopted when calculating physical conditions from line flux ratios of metal ions (see \S\ref{s:tecalcs}). The reddening corrected line intensities relative to H$\beta$ in the full LBT \yp\ Project sample are reported in Table \ref{t:allinten}.

\subsection{Emission Line Diagrams}\label{s:diagrams}

To assess the ionization conditions of the sample, we plot commonly utilized emission line ratios measured in the rest-frame optical/NIR in Figures \ref{fig:bptfigs} and \ref{fig:extra_bptfigs}. Figure \ref{fig:bptfigs} plots the traditional diagnostics proposed by \citet[][or BPT]{bald1981} and \citet{veil1987}, or \oiii/H$\beta$ vs.\ \nii/H$\alpha$, \sii/H$\alpha$, and \oi/H$\alpha$. The dashed black and dotted green lines denote the \citet{kauf2003} and \citet{kewl2001} demarcations for ionization from an AGN and extreme starburst galaxies, respectively. Ionized regions with line ratios above these demarcations likely have ionization sources beyond simple stellar populations, such as AGN or shocks. The line ratios measured in the LBT \yp\ Project nebulae, plotted as blue diamonds, are generally offset to high \oiii/H$\beta$ and low \nii/H$\alpha$, a pattern that is typical for high-ionization, low-metallicity nebulae \citep{kewl2013}. Numerous galaxies have log(\oiii/H$\beta$)$\sim$0.3 at log(\nii/H$\alpha$)$<$$-$2, indicative of very low gas-phase metallicities that result in the low intensity of \oiii.

For comparison, we include the emission line ratios measured in extragalactic \hii\ regions observed as part of the CHemical Abundances Of Spirals project \citep[CHAOS,][]{berg2015}. We use the emission line intensities reported in \citet{berg2020}, \citet{roge2021}, and \citet{roge2022}. We select only \hii\ regions with significant emission line detections for all relevant lines in each diagram. Purple circles in Figure \ref{fig:bptfigs} highlight the line ratios in the CHAOS \hii\ regions, which are consistent with the trends expected for more metal-rich nebulae and softer ionizing spectra.

The CHAOS \hii\ regions occupy clearly distinct areas of the N2-BPT diagram from the LBT \yp\ Project sample, but there is some overlap between the two samples in the S2- and O1-BPT diagrams (center and right panels of Figure \ref{fig:bptfigs}). Additionally, some nebulae fall along the \citet{kewl2001} maximum starburst demarcation in these same diagrams, a trend that could be produced by other ionization sources in the photodissociation region (PDR) that contains low-ionization species such as O$^0$ and S$^+$. Indeed, a small number of CHAOS \hii\ regions also scatter above the line ratios expected for a maximum starburst, a trend that has been associated with shock ionization \citep{crox2015}. Shock ionization from supernovae or Wolf-Rayet (WR) stars can bias the measured gas-phase physical conditions and chemical abundances from an emission line spectrum \citep{skil1985}. However, shock contribution is difficult to assess from the integrated spectra of the LBT \yp\ Project nebulae beyond general emission line diagrams or the detection of emission lines from very high ionization ions like Ne$^{4+}$ \citep{izot2021}. While some nebulae may have significant shock contributions to these high-ionization lines \citep[such as SHOC133, also reported by][]{izot2021}, we do not attempt to correct for the shock contribution owing to the general agreement with the star-forming areas in the emission line diagrams of Figure \ref{fig:bptfigs}.

These ionization conditions are further explored in the panels of Figure \ref{fig:extra_bptfigs}. In the left panel, we plot O$_{32}$ vs.\ R$_{23}$, where O$_{32} = $ I(\oiii$\lambda$4959,5007)/I(\oii$\lambda$3727) and R$_{23} = $ (I(\oiii$\lambda$4959,5007) + I(\oii$\lambda$3727))/I(H$\beta$). While some low-metallicity nebulae have similar R$_{23}$ as the high-metallicity \hii\ regions, the majority of the sample contains more highly ionized ISM and is offset to high R$_{23}$. The area of parameter space covered by the LBT \yp\ Project targets in the ionization-excitation diagram is similar to the average conditions observed in high-\emph{z} star-forming galaxies, namely log(O$_{32}$) $>$ 0 and log(R$_{23}$) $>$ 0.5 \citep{shap2015,shapley2025,stro2017}.

A similar trend is found when plotting log(I(\neiii$\lambda$3869)/I(\oii$\lambda$3727)) vs.\ log(I(\neiii) + I(\oii)/I(H$\delta$)) in the center panel. This diagnostic diagram has grown in popularity owing to the rest-optical observations of high-\emph{z} star-forming galaxies with JWST \citep{came2023,robe2024}. Similar to O$_{32}$, log(I(\neiii$\lambda$3869)/I(\oii$\lambda$3727)) traces the ionization state of the nebula via the difference in the IPs of Ne$^{2+}$ and O$^+$. The sum of \neiii\ and \oii\ relative to a \hi\ recombination line is sensitive to metallicity and, therefore, is akin to the R$_{23}$ parameter. The low-metallicity nebulae generally have log(I(\neiii) + I(\oii)/I(H$\delta$)) $<$ 0.8, consistent with metal-poor ISM ionized by a hard ionizing spectra. Some LBT \yp\ Project targets have similar \neiii/\oii\ intensity ratios as those observed in other high-\emph{z} EELGs \citep[e.g.,][]{came2023,bunk2023,topp2024,shapley2025}.

We further illustrate the presence of very high ionization conditions in the LBT \yp\ Project nebulae in the right panel of Figure \ref{fig:extra_bptfigs}. In this panel, we plot the intensity of \heii$\lambda$4686 relative to H$\beta$ against the \nii/H$\alpha$ intensity ratio. When reporting \heii/H$\beta$, we only consider the narrow component of \heii\ with FWHM similar to other neighboring emission lines (\feiii$\lambda$4658, \ariv$\lambda$4740, etc.) and omit any broad \heii\ emission. However, the \heii\ intensity measured in the CHAOS \hii\ regions may include a portion of broad \heii\ flux from, for example, WR stars \citep{lope2010}. The red dot-dashed line from \citet{shir2012} denotes line ratio predictions when 10\% of the total line flux is produced by AGN ionization. All LBT \yp\ Project targets fall below this demarcation and are consistent with predominant stellar ionization. A total of 50 nebulae exhibit significant narrow \heii\ emission, ranging from $\sim$0.2\% to more than 3\% the intensity of H$\beta$; however, the ionizing spectrum of simple stellar populations, even at low metallicity or when considering binary interactions, predict very few He$^+$-ionizing photons \citep{eldr2017}. The detection of \heii\ in the LBT \yp\ Project sample adds to an ongoing problem in nebular astrophysics, where the observed \heii\ emission is much larger than predicted from ionization by a stellar population alone \citep[see discussion in][]{garn1991,senc2017,berg2021}. Other sources such as shocks, X-ray binaries, evolved stellar populations (i.e., WR stars, stripped stars), or AGN contamination could contribute to the production of very high-energy photons and the observed nebular \heii\ emission \citep{gotb2019,senc2020,flury2025,hovis-afflerbach2025}. Further observations in different spectroscopic bands \citep[UV, MIR, and NIR, see][hereafter Paper III]{Weller2026} are required to further test the ionization structure and sources of these metal-poor nebulae, particularly those with similar emission line characteristics as those observed in high-\emph{z} EELGs.

\section{Physical Conditions}\label{s:phys}

\subsection{\texorpdfstring{n$_e$}{ne} and \texorpdfstring{T$_e$}{Te} Calculations}\label{s:tecalcs}

We now measure the physical conditions of the ionized gas, namely \te\ and \den, in the LBT \yp\ Project sample using numerous emission lines from different metal ions. To do so, we use the \textsc{PyNeb} package \citep{luri2012,luri2015L} version 1.1.18 and the atomic data listed in Table \ref{t:atomic}. We assume a three-zone ionization structure in each nebula, where each zone is composed of ions with similar IP and is described by a characteristic \te. The low-ionization zone contains ions with IP $\lesssim$ 23 eV, such as N$^+$, O$^+$, and S$^+$. The ions S$^{2+}$, Cl$^{2+}$, and Ar$^{2+}$ have 23 $\lesssim$ IP $\lesssim$ 35 eV and make up the intermediate-ionization zone. Finally, we assume the high-ionization zone contains ions with IP $>$ 35 eV: O$^{2+}$ and Ar$^{3+}$. We note that many other ions are detected in the MODS spectra, but are not used in the present analysis of the gas-phase physical conditions (Rogers et al., in Prep, hereafter Paper VI). In the LBT \yp\ Project, the detection rate of narrow, nebular \heii\ is substantial: 50 objects have \heii$\lambda$4686 detections at S/N greater than 3 (see right panel in Figure \ref{fig:extra_bptfigs} and Table \ref{t:hiinten}). While a fourth, very high ionization zone may be preferred for ions with IP $>$ 54 eV \citep[see][]{berg2021} and characterized by a \te\ measurement from the \neiii\ emission lines, the \neiii$\lambda$3342 auroral line is not detected in a significant fraction of the sample to warrant the use of a different electron temperature.

\begin{deluxetable}{lcc}  
\tablecaption{Selected Atomic Data \label{t:atomic}}
\renewcommand{\arraystretch}{1.0}
\tablewidth{\textwidth}
\tabletypesize{\footnotesize}
\tablehead{
   \colhead{} & 
   \colhead{Transition} & 
   \colhead{Collision} \\ [-0.2cm]
   \colhead{Ion}  & 
   \colhead{Probabilities} &
   \colhead{Strengths}}
\startdata
N$^+$ 	& \citet{froe2004}  &  \citet{taya2011}  \\ 
\hline
O$^+$ 	& \citet{froe2004}  &  \citet{kisi2009} \\
\hline
O$^{2+}$ 	& \citet{froe2004}  &   \citet{stor2014} \\
\hline
S$^+$   & \citet{irim2005}  & \citet{taya2010} \\
\hline
S$^{2+}$    & \citet{froe2006}  &   \citet{huds2012}   \\
\hline
Cl$^{2+}$    & \citet{rynk2019}  &   \citet{butl1989}   \\
\hline
Ar$^{2+}$   &   \multrow{\citet{mend1983b}, \\ \citet{kauf1986}}    &   \citet{gala1995}  \\
\hline
Ar$^{3+}$   &   \citet{mend1982}    &   \citet{rams1997} \\
\hline
\enddata
\tablecomments{The atomic data selected for the CELs utilized for physical conditions and O/H abundances.}
\end{deluxetable}

Although CEL fluxes are sensitive to ionic abundance, \te, and \den, ratios of two CELs originating from the same ion eliminates the dependence on the abundance and is only sensitive to the gas-phase physical conditions. If the two emission lines are produced by energy levels of similar excitation energy but different critical densities, then the ratio is primarily sensitive to \den. Density-sensitive line ratios that are routinely detected in the LBT \yp\ Project sample are \sii$\lambda$6717/\sii$\lambda$6731, \oii$\lambda$3729/\oii$\lambda$3726, \cliii$\lambda$5518/\cliii$\lambda$5538, and \ariv$\lambda$4711/\ariv$\lambda$4740. The \oii$\lambda$3726,29 strong lines and \ariv$\lambda$4711+\hei$\lambda$4713 are partially blended at the resolution of MODS, but the line fitting approach discussed in \S\ref{s:linefit} allows for a constraint on \den\ from these ions. Notably, the wavelength separation of the \den-sensitive emission lines is sufficiently small such that the flux ratio is insensitive to $E(B-V)$. Therefore, we use the line flux ratios of the CELs without additionally correcting for reddening.

To measure \den\oii\ and \den\sii, we apply the \textsc{getCrossTemDen} function along with the measured \sii\ and \oii\ line ratios and the \nii$\lambda$5755/\nii$\lambda$6584 \te-sensitive ratio to solve for the combination of \te\ and \den\ required to produce the observed line ratios. If the \te-sensitive \nii\ line ratio cannot be constrained, then we use the \oiii$\lambda$4363/\oiii$\lambda$5007 ratio instead, but this temperature does not reflect the low-ionization gas containing \oii\ and \sii. Instead, we use this estimated \te\oiii\ to infer the low-ionization zone \te, then recompute \den\oii\ and \den\sii\ at the inferred temperature. When measuring \den\cliii\ and \den\ariv, we use the \textsc{getTemDen} function at the measured intermediate- and high-ionization zone \te\ from \siii\ and \oiii, respectively. To estimate the uncertainty on \den, we generate a normal distribution of 2000 line ratios with center and standard deviation equal to the measured line ratio and its uncertainty, respectively. Ratios not permitted by the atomic data are excluded, then density is recomputed from the ensemble. The standard deviation in the ensemble is used to measure the asymmetric uncertainties on \den.

With direct constraints on \den\ in multiple ionization zones, we adopt a characteristic density in the ISM that will be used for subsequent \te\ and abundance calculations. Following the methods of \citet{mend2023}, we prioritize different density diagnostics based on the measured \den\sii. When \den\sii\ $<$ 100 cm$^{-3}$, the ISM is in the low-density limit, where \te\ and abundance calculations using optical emission lines are insensitive to \den, so we perform all calculations at \den\ $=$ 100$\pm$50 cm$^{-3}$. If 100 cm$^{-3}$ $<$ \den\sii\ $<$ 1000 cm$^{-3}$, we use an average of \den\sii\ and \den\oii, or \den\sii\ when the latter is unconstrained. \citet{mend2023} recommend using a combination of \den\ measurements from \sii, \oii, \feiii, \cliii, and \ariv\ when \den\sii\ $>$ 1000 cm$^{-3}$, although these high \den\sii\ are not measured in the LBT \yp\ Project nebulae.

A measurement of \te\ requires two CELs originating from the same ion but with a large separation in excitation energy. The following \te-sensitive intensity ratios are possible to measure from a MODS spectrum: \nii$\lambda$5755/\nii$\lambda$6584, \oii$\lambda$7320,30/\oii$\lambda$3726,29, \oiii$\lambda$4363/\oiii$\lambda$5007, \neiii$\lambda$3342/\neiii$\lambda$3869, \sii$\lambda$4069,76/\sii$\lambda$6717,31, \siii$\lambda$6312/\siii$\lambda$9069,9532, and \ariii$\lambda$5192/\ariii$\lambda$7135. To measure \te\ from one of these line ratios, we use the \textsc{PyNeb} \textsc{getTemDen} function, the characteristic \den\ discussed above, and the measured flux ratio of the emission lines, corrected for reddening using the $E(B-V)$ determined from \S\ref{s:redcorr}. When \den\sii\ and \den\oii\ are available, we use these densities when determining \te\sii\ and \te\oii, respectively, for a consistent temperature measurement in the \sii\ and \oii\ emitting gas. The uncertainty on $E(B-V)$ is propagated into the line flux ratio, which is dependent on the wavelength separation of the auroral and nebular emission lines; for example, the ratio of \oiii\ emission lines is less affected by a poorly-constrained $E(B-V)$ than the ratio of the \oii\ auroral and nebular lines. Similar to density uncertainties, we generate an ensemble of 1000 line ratios using a normal distribution with center and standard deviation equal to the measured line ratio and its uncertainty, respectively. We repeat the \te\ calculation for the full distribution and take its standard deviation to be the uncertainty in the measured \te.

This general procedure is applied with a few notable caveats. First, the \oiii$\lambda$4363 emission line can be contaminated by the neighboring \feii$\lambda$4359 \citep{andr2013,curt2017,khoram2025}. The contaminating \feii\ line, if not properly accounted for, will bias the inferred \te\oiii\ high and the O$^{2+}$ abundance low, particularly in regions of high-metallicity where \feii$\lambda$4359 has been detected at comparable strengths to \oiii$\lambda$4363 \citep{roge2022}. However, one may not expect significant \feii\ contamination given the observational characteristics of the average nebula in this sample. For instance, the LBT \yp\ Project nebulae are selected for their low gas-phase O/H, where the abundance of Fe is expected to be relatively low \citep[roughly a factor of 55$\times$ less than the gas-phase O abundance, see][]{mend2024}. Further, while dust depletion on Fe should be minimal in the high-ionization ISM, what Fe is present will be in higher ionization states than Fe$^+$ (see \S\ref{s:diagrams}). Fluorescence pumping is another important excitation mechanism for the $^6S_{5/2}$ level which produces \feii$\lambda$4359, potentially decoupling the CEL intensity from the Fe$^+$ abundance \citep{mendoza2023}.

We detect the isolated \feii$\lambda$4287 emission line in 13 nebulae. This emission line is produced by the same upper energy level as \feii$\lambda$4359, such that the intensity ratio of the two lines is fixed at I(\feii$\lambda$4359)/I(\feii$\lambda$4287) $=$ 0.73 \citep[using the atomic data of][]{baut2015}. For nebulae with significant \feii$\lambda$4287 detections, we subtract the inferred \feii$\lambda$4359 intensity from \oiii$\lambda$4363 before calculating \te\oiii. For nebulae without significant \feii$\lambda$4287 emission, the contamination of the \oiii\ auroral line is assumed to be negligible. The maximum magnitude of \feii\ contamination is measured in HS1222+3741 and I Zw 18 SE, where the inferred \feii$\lambda$4359 results in a 3.3\% reduction in the measured \oiii$\lambda$4363 intensity. For the 13 nebulae, the median \feii\ contamination accounts for just 1.1\% of the \oiii\ intensity; therefore, \feii\ contamination is often not a major consideration in these low-metallicity, high-ionization environments.

The second major caveat concerns contamination of the rest-NIR emission lines from the night sky. The redshift of the sample is broad, ranging from $-$0.00034 to 0.09193, and these observations were conducted over many different nights with varying observing conditions (see Paper I). As such, emission lines of interest can redshift into the atmospheric O$_2$ A and B bands and areas of telluric absorption \citep{noll2012}.
While it is difficult to assess the contamination of individual emission lines, contamination of the rest-NIR \siii\ strong lines can be assessed using the fixed intensity ratio of the two lines, I(\siii$\lambda$9532)/I(\siii$\lambda$9069) $=$ 2.47, and the position of known telluric features.
Before calculating \te\siii, we first measure the intensity ratio of the \siii\ strong lines and compare to the theoretical ratio. When the ratio agrees with the theoretical ratio, we use the sum of the two strong lines to calculate \te\siii.
If the ratio significantly deviates from 2.47, then telluric absorption is potentially reducing the intensity of one of the strong lines. For example, when the measured ratio is less than 2.47 we prioritize \siii$\lambda$9069 for \te\siii\ unless it is coincident with the position of strong telluric absorption, at which point we use \siii$\lambda$9532. However, if the measured \siii\ ratio deviates by $>$3$\sigma$ below theory, then \siii$\lambda$9532 is significantly contaminated and we use \siii$\lambda$9069 irregardless of the position of the telluric features. A similar procedure is taken for line ratios greater than theory, where \siii$\lambda$9532 is, generally, prioritized.
Of the 58 nebulae with both \siii\ lines detected, 27 have \siii$\lambda$9532/\siii$\lambda$9069 ratios that agree with theory, 18 have ratios less than theoretical, and the remaining 13 exhibit ratios greater than theoretical. If only one \siii\ line is detected, then we use that emission line for \te\siii\ calculation. The \te\ and \den\ for the LBT \yp\ Project sample are reported in Table \ref{t:abun}.

\begin{figure*}[t]
   \centering
   \includegraphics[width=0.99\textwidth]{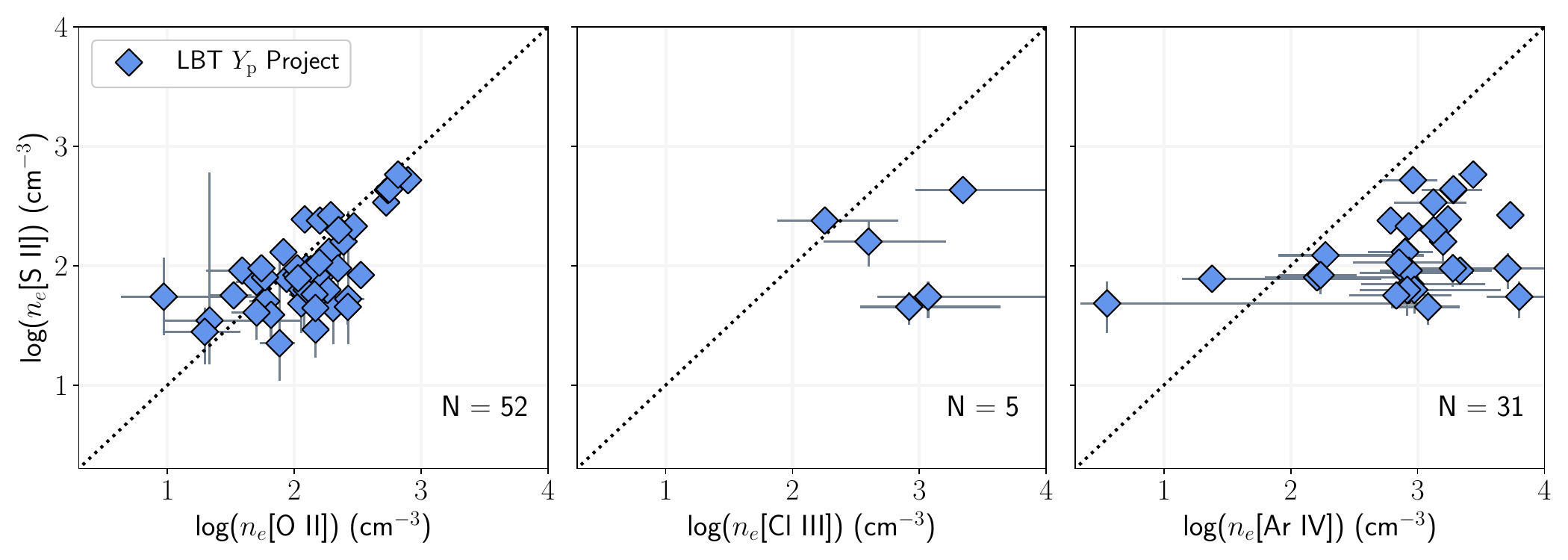}
   \caption{Direct density trends measured in the LBT \yp\ Project sample. \den\sii\ is plotted against \den\oii\ in the low-ionization zone (Left), \den\cliii\ in the intermediate-ionization zone (Center), and \den\ariv\ in the high-ionization zone (Right). The dotted black line represents equivalent densities. The number of nebulae with simultaneous \den\ is provided in the bottom right of each panel. The \sii\ and \oii\ densities scatter around the one-to-one line, but \den\cliii\ and \den\ariv\ are systematically higher than the low-ionization zone densities.}
   \label{fig:ne-ne}
\end{figure*}

\subsection{\texorpdfstring{n$_e$}{ne} Trends}\label{s:dentrends}

Given the direct constraints on various ISM properties in the LBT \yp\ Project nebulae, it is worthwhile to investigate how physical conditions in different ionization zones scale with one another. In Figure \ref{fig:ne-ne}, we compare the simultaneous \den\ measured from four different ions, where the number of galaxies is provided in the bottom right of each panel. The most common \den\ measurements are from the \oii\ and \sii\ strong lines, and the two densities plotted in the left panel show good agreement with non-negligible scatter. The Spearman correlation coefficient between \den\sii\ and \den\oii\ is $r=$ 0.54 with a $p$ value of 3$\times$10$^{-5}$, indicating that the density measurements from the O$^+$ and S$^+$ ionization states are moderately correlated. Similar consistency between \den\oii\ and \den\sii\ has been reported in other samples of ionized nebulae \citep[e.g.,][]{mend2023}, and this trend may be expected if both ions reside in the same volume of the \hii\ region.

Defining the difference between the simultaneous \den\oii\ and \den\sii\ in the left panel as $\delta$\den, we find the average and standard deviation of $\delta$\den\ are 56 cm$^{-3}$ and 81 cm$^{-3}$, respectively. The average error on \den\sii\ and \den\oii\ is $\lesssim$ 30 cm$^{-3}$, indicating that the tendency to lower \den\sii\ in the ISM and large scatter are significant. The lower \den\sii\ could be related to a bias introduced to the \sii\ lines by Diffuse Ionized Gas (DIG) in the integrated spectrum of a galaxy's ISM.
Photons with energies between 10.4 eV and 13.6 eV are sufficient to ionize S atoms, but are incapable of ionizing H or O in the ISM. Therefore, \sii\ emission may originate in gas both outside and within the \hii\ region, and the DIG has lower electron density than typically observed in \hii\ regions \citep{reyn1991,berk2008}. This bias may be an important consideration in more distant LBT \yp\ Project nebulae where the MODS longslit captures the central star-forming region and the DIG, resulting in increased scatter to lower \den\sii. Alternatively, a highly stratified ISM could result in large variations between \den\ measured in the different ionization zones. For example, the ISM containing O$^+$ also contains S$^+$, S$^{2+}$, Cl$^{2+}$, and Ar$^{2+}$ owing to the high IP of O$^+$. Should the electron density be larger near the ionizing sources and in the ISM with higher-ionization ions, then the \den\oii-\den\sii\ trends may be the result of a stratified ISM.

This can also be assessed with density measurements from other high-ionization ions in the LBT \yp\ Project nebulae. The next two panels of Figure \ref{fig:ne-ne} plot \den\sii\ against the intermediate- and high-ionization zone densities from \cliii\ and \ariv, respectively. Owing to the faint lines and the increased noise near the dichroic crossover in the MODS detectors at $\sim$5700 \AA, there are few nebulae where \den\cliii\ can be reliably constrained. However, the CELs of \sii\ and \ariv\ are significantly detected in 31 nebulae, allowing for a comparison of the low- and high-ionization zone densities. The average \den\ariv\ plotted in the right panel of Figure \ref{fig:ne-ne} is $\sim$1500 cm$^{-3}$, showing a clear trend of increasing \den\ with ionization state and corroborating density trends reported in other EELGs from optical \citep{berg2021,mend2023} and UV CELs \citep{ming2022,topping2025}.

An important caveat to these findings is that the \sii\ and \ariv\ line intensity ratios are sensitive to \den\ in different regimes. The critical density of \sii$\lambda$6717 at 1.75$\times$10$^4$~K is 1500 cm$^{-3}$, such that the \sii\ line ratio is most sensitive to density above 10$^2$ cm$^{-3}$ and below $\sim$1.5$\times$10$^3$ cm$^{-3}$. The \ariv\ line ratio is sensitive to \den\ between $\sim$10$^{3}$ and 10$^{5}$ cm$^{-3}$. The difference in critical densities may contribute to the offset observed in the \den\sii-\den\ariv\ panel of Figure \ref{fig:ne-ne}. This can be assessed using the auroral lines of \sii\ to recompute \den; as \citet{mend2023} discuss, the \sii\ and \oii\ auroral-to-nebular line ratios are sensitive to \den\ up to $\sim$10$^5$ cm$^{-3}$ due to the high critical densities of the energy levels that produce auroral \oii\ and \sii\ emission. To assess if higher \den\ is masked by the nebular \oii\ and \sii\ ratios, we redetermine \den\oii\ and \den\sii\ using the \oii$\lambda$7320,30/\oii$\lambda$3727 and \sii$\lambda$4069,71/\sii$\lambda$6717,31 ratios, respectively. This requires an independent measure of \te, so we assume a fixed low-ionization zone \te\ (either the measured \te\nii\ or using a \te\ scaling relation, see discussion in \S\ref{s:dirabun}).

\begin{figure}[t]
   \centering
   \includegraphics[width=0.47\textwidth]{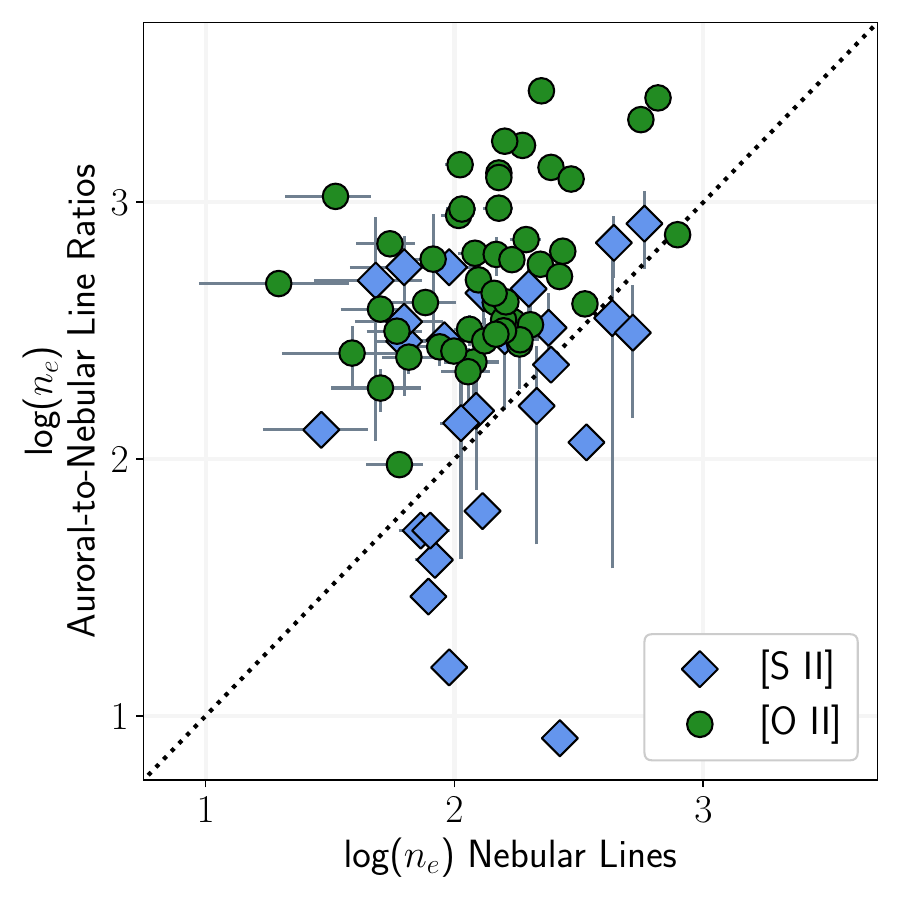}
   \caption{Electron density measured in the LBT \yp\ Project nebulae from the \oii\ (green circles) and \sii\ emission lines (blue diamonds). The horizontal axis represents \den\ measured from the nebular line ratios \oii$\lambda$3729/\oii$\lambda$3726 and \sii$\lambda$6731/\sii$\lambda$6717. The vertical axis represents \den\ measured from the auroral-to-nebular line ratios \oii$\lambda$7320,30/\oii$\lambda$3727 and \sii$\lambda$4069,76/\sii$\lambda$6717,31 assuming a fixed low-ionization zone \te. In general, the \oii\ and \sii\ auroral line ratios suggest larger \den\ than the nebular lines alone. The higher densities lead to increased collisional de-excitation of the \oii\ and \sii\ nebular lines which can, potentially, bias electron temperature and ionic abundance calculations.}
   \label{fig:ds2do2}
\end{figure}

In Figure \ref{fig:ds2do2}, the resulting auroral-to-nebular \den\ are plotted against the standard \den\ calculated from the nebular lines only. The blue diamonds and green circles are calculated using the line ratios from \sii\ and \oii, respectively. Consistent with \citet{mend2023}, we find that the auroral-to-nebular \oii\ ratios indicate values of \den\ moderately higher than the nebular lines alone, assuming the characteristic low-ionization zone \te\ is reflective of the gas containing O$^+$ and S$^+$. The \sii\ auroral-to-nebular line ratios produce large scatter in \den, which may be the result of a different \te\ in the S$^+$ gas that deviates from the assumed low-ionization zone \te. The average \den\oii\ and \den\sii\ increase to 705$\pm$84 cm$^{-3}$ and 276$\pm$42 cm$^{-3}$, respectively, which is larger than implied from the nebular lines alone but still less than the average \ariv\ densities.

Significant density stratification could systematically bias the \te\ and ionic abundances, which assume the characteristic ISM density discussed in the previous section. Of upmost importance is collisional de-excitation of \oiii$\lambda$5007: O$^{2+}$ is the most abundant ionization state in the high-ionization ISM, and emission from \oiii$\lambda$5007 effectively regulates the cooling structure in the metal-poor gas \citep{osterbrock2006,ryden2021}. As such, \te\oiii\ measured from the CELs of \oiii\ is the most relevant temperature for O$^{2+}$/H$^+$ and the total O abundance in the ISM of the LBT \yp\ Project nebulae. \oiii$\lambda$5007 originates from an upper energy level with critical density of $\sim$8$\times$10$^5$ cm$^{-3}$ at \te\ $=$ 1.75$\times$10$^4$ K. If \den\ariv\ is appropriate for the high-ionization zone and the marginal collisional de-excitation in \oiii$\lambda$5007 is not considered, then the measured \te\oiii\ and O$^{2+}$/H$^+$ could be biased significantly high and low, respectively \citep{martinez2025}. 
The critical density of the energy levels that produce \oii$\lambda$3726,29 are on the order of 10$^3$ cm$^{-3}$, while \nii$\lambda$6584's critical density is $\sim$1.1$\times$10$^5$ cm$^{-3}$ at \te\ $=$ 1.75$\times$10$^4$ K. A similar bias to high \te\oii\ may be present if the low-ionization zone density is at or above the critical density of \oii$\lambda$3726,29 (see discussion in \S\ref{s:tlow}). \te\nii\ is less sensitive to high densities, but the larger \den\ in Figure \ref{fig:ds2do2} measured from the auroral-to-nebular line ratios of \sii\ and \oii\ will increase the ionic abundance of O$^+$. The net result of higher densities in the low- and high-ionization zones is a potentially larger O/H abundance that could alter the He/H vs.\ O/H relation.

To assess whether these higher densities should be considered, we repeat the \te\oiii\ and O$^{2+}$ ionic abundance calculations using \den\ariv\ when available, then recompute O$^+$ using the higher characteristic \den\ inferred from the \oii\ and \sii\ auroral-to-CEL ratio and the low-ionization zone \te.
In all but one case\footnote{The one galaxy where the change in O/H is significant is WJ0851$+$5841, where an increase in the adopted \den\ by more than a factor of 8 and 5 in the low- and high-ionization zones, respectively, produces a 7\% larger O/H.}, adopting the higher ISM densities produces a fractional change to the total O/H abundance that is less than the average uncertainty in O/H ($\sim$4\% for the full LBT \yp\ Project sample). 
Therefore, considering higher \den\ does not produce a meaningful change in the total O/H in the LBT \yp\ Project nebulae, and we proceed with the characteristic \den\ discussed in \S\ref{s:tecalcs}.
Only direct measurements of \den\ in the different ionization zones can assess the presence of very high-density gas, which will strongly affect chemical abundance measurements in high-z EELGs with \den\ in excess of 10$^6$ cm$^{-3}$ \citep[e.g.,][]{topp2024,haye2025}.

Significant density stratification can also affect the measurement of He/H. The intensity of \ion{He}{1} recombination lines are dependent on \te, \den, and He$^+$ abundance, but the $2^3S$ level is metastable. As a result, \ion{He}{1} triplet lines are also sensitive to radiative transfer effects, where high-energy \ion{He}{1} lines like \ion{He}{1} $\lambda$3889 are absorbed and re-emitted as redder lines like \ion{He}{1} $\lambda$7065 \citep{benjamin2002,kurichin2026}. 
This trend can be degenerate with density.  This is why we solve simultaneously for all parameters, including density, optical depth, and He$^+$/H$^+$, based on many He and H emission lines. We then employ a MCMC to assess the increased uncertainties on these parameter determinations due to degeneracies (see Papers I and IV). While the intensity of individual \ion{He}{1} lines can vary significantly with temperature, density, and optical depth, the inclusion of \ion{He}{1} $\lambda$10830 greatly reduces this degeneracy owing to its strong density dependence \citep[][see also Paper I]{izot2014,aver2015}. This motivated the use of LBT/LUCI to observe this crucial emission line (Paper III), and we refer the reader to Paper IV for an overview of the radiative transfer modeling used in the LBT \yp\ Project to simultaneously fit for these physical conditions and the He abundance.   To be clear, although densities derived from CELs are reported in this paper, they are not used in the derivation of the He abundances.

\section{Temperature Scaling Relations}\label{s:tescaling}

With the wealth of \te\ information available from this sample of ionized regions, we now provide empirical constraints on the trends of \te\ as measured from different ions within the same nebula. These \te\ scaling relations, often referred to as \te-\te\ relations, are crucial for typical direct abundance studies: when an ionization zone temperature is missing, these \te\ scaling relations can be used to infer the missing \te\ for chemical abundance calculations in the ionization zone of interest. \te\ scaling relations have been calibrated via two methods, the first of which utilizes empirical direct \te\ data from large samples of individual ionized regions \citep{este2009,crox2016,yate2020,roge2021,mend2023,scholte2026} or from spaxels of IFU observations in star-forming galaxies \citep{rick2024,khoram2025}. Obtaining a statistically-significant sample of direct \te\ measurements from multiple ions is observationally challenging, requiring a combination of high sensitivity and broad wavelength coverage to detect the auroral emission lines and their strong line counterparts, respectively. The second method of calibrating \te\ scaling relations circumvents these problems by utilizing photoionization models to predict the trend of \te\ in different ionization zones over a range of model ionization parameters and metallicities \citep{camp1986,garn1992,page1992,izot2006}. 
The functional form of the \te\ scaling relations has a profound impact on chemical abundance measurements when direct \te\ data are missing, and the ionized nebulae in this study present an opportunity to recalibrate these relations at high \te/low metallicity.

\subsection{\texorpdfstring{T$_e$\siii}{TeS3}-\texorpdfstring{T$_e$\oiii}{TeO3}}

\begin{figure}[t]
   \centering
   \includegraphics[width=0.47\textwidth]{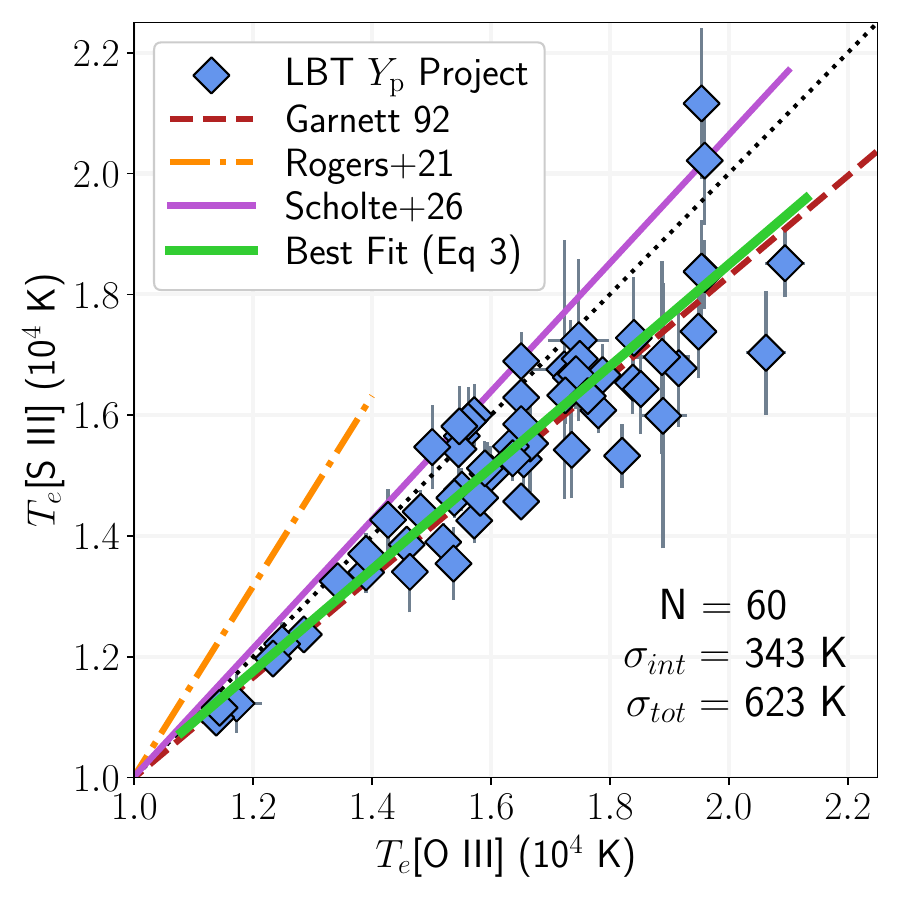}
   \caption{Simultaneous \te\siii\ and \te\oiii\ measured from 60 LBT \yp\ Project targets are plotted in blue. The black dotted line represents equivalent temperatures. The red dashed line is the photoionization model \te\ scaling relation from \citet{garn1992}, the empirical scaling relations derived by \citet{roge2021} and \citet{scholte2026} are provided as the dot-dashed orange and solid purple lines, respectively. The best-fit scaling relation to the low-metallicity nebulae (green solid line) is similar to the photoionization model \te\ scaling relation. While small compared to the uncertainties on \te\siii, the intrinsic and total dispersion in the empirical \te\ about this relation (provided in the bottom right) are non-negligible.}
   \label{fig:ts3to3}
\end{figure}

\te\oiii\ is measured in every target, and the next most common \te\ measurement is made from \siii$\lambda$6312. Only two galaxies, AGC198691 and HSCJ2314+0154, lack a significant detection of the \siii\ auroral line. The low excitation energy of the \siii\ $^1$S$_0$ level, 3.37 eV, relative to other commonly-utilized auroral lines (e.g., \oiii$\lambda$4363 at 5.35 eV) makes this line relatively easy to detect in both low- and high-metallicity nebulae \citep[e.g.,][]{crox2016}. If, empirically, \te\siii\ and \te\oiii\ trace one another over a broad range in \te, then the \siii\ auroral line may offer a reliable method to infer the temperature of the \oiii\ emitting region when \oiii$\lambda$4363 is not detected. In Figure \ref{fig:ts3to3}, we plot the \te\oiii-\te\siii\ trends measured in 60 low-metallicity nebulae as blue diamonds. The black dotted line represents equivalent \te\ as measured from the two ions. There is a tight correlation between the intermediate- and high-ionization zone temperatures at 1.1$\times$10$^4$ K $<$ \te\oiii\ $<$ 2.1$\times$10$^4$ K, indicating that these two \te\ are related to one another in the ISM.

Three \te\ scaling relations from the literature are plotted in Figure \ref{fig:ts3to3}. The \citet{garn1992} photoionization model relation (dashed red) is a commonly adopted scaling relation used to infer the intermediate-ionization zone \te\ when a direct \te\siii\ is unavailable. The \citet{roge2021} empirical relation (dot dashed orange) is plotted over the range of \te\ observed in the relatively metal-rich CHAOS \hii\ regions. To appreciate the behavior of the \te\siii-\te\oiii\ at high temperature, we also include the \citet{scholte2026} relation (solid purple). This relation is measured from 3781 galaxies in the Dark Energy Spectroscopic Instrument \citep[DESI,][]{desi2024} survey, including a significant number of galaxies with \te\oiii\ $>$ 1.5$\times$10$^4$ K. These high-\te\ are most comparable to the temperatures trends measured in the LBT \yp\ Project, although these nebulae deviate from the DESI relation towards higher \te\oiii\ at fixed \te\siii. This may be related to the larger \te\ range considered in the DESI sample and a change in slope of the \te\ scaling relation (qualitatively supported by the CHAOS \hii\ regions, see below).

With the large number of homogeneous \te, we derive a new \te\siii-\te\oiii\ scaling relation using the approach of \citet{roge2021}. In brief, we use the \textsc{scipy} implementation of Orthogonal Distance Regression (ODR) to fit a linear relation to the \te\ data considering the uncertainties on both \te\siii\ and \te\oiii. We then use the best-fit parameters from the ODR fit in the \textsc{python linmix} package\footnote{\url{https://github.com/jmeyers314/linmix}}, which is based on the IDL fitting routine of \citet{kell2007}, to determine the intrinsic random dispersion, $\sigma_{int}$, in the dependent variable about the line of best fit. The total scatter in \te, $\sigma_{tot}$, about the relation is determined using the formulation outlined in \citet{bedr2006}, and this is provided in the lower right of Figure \ref{fig:ts3to3}. We repeat this process for the inverse relation to determine the random and total dispersion in the other \te, which is not necessarily the same due to the asymmetric uncertainties on \te\siii\ and \te\oiii\ \citep[see discussion in][]{roge2021}. The best-fit relation for the \te\siii-\te\oiii\ trend is plotted in Figure \ref{fig:ts3to3} as a green solid line and has the functional form
\begin{equation}\label{eq:ts3o3}
  \begin{gathered}
    T_{e}\mbox{[S III]}=0.84(\pm0.04)\times T_{e}\mbox{[O III]} + 1700(\pm600) \ \mbox{K} \\
    \sigma_{int} = 340 \pm 90 \ \mbox{K}.
  \end{gathered}
\end{equation}
The inverse relation, which could be used to infer \te\oiii\ from a measured \te\siii, is described by
\begin{equation}\label{eq:to3s3}
  \begin{gathered}
    T_{e}\mbox{[O III]}=1.19(\pm0.06)\times T_{e}\mbox{[S III]} - 2000(\pm800) \ \mbox{K} \\
    \sigma_{int} = 420 \pm 110 \ \mbox{K}.
  \end{gathered}
\end{equation}
When applying these \te\ relations to infer a missing \te, we recommend adding $\sigma_{int}$ in quadrature to the propagated uncertainty of the measured \te\ \citep[see Eq.\ 8 in][]{roge2021}. This approach reflects the limitations of the linear \te\ scaling relations to match the empirical, random scatter in the temperature trends of ionized nebulae.

The empirical \te\ data and scaling relations of Eqs \ref{eq:ts3o3} and \ref{eq:to3s3} agree remarkably well with the photoionization model predictions of \citet{garn1992}, consistent with findings from other direct abundance studies \citep[e.g.,][]{kenn2003b,mira2023}. However, Eq \ref{eq:ts3o3} diverges from the empirical measurements made in \citet{roge2021} and \citet{scholte2026}, both of which find that a steeper slope describes the \te\siii-\te\oiii\ scaling relation. The majority of the \hii\ regions used in calibrating the \citet{roge2021} empirical relations have \te\ $\lesssim$ 10$^4$ K, indicating that a different functional form may be necessary for more metal-rich/cooler nebulae. This is qualitatively supported by the DESI data, where the highest density of galaxies in \citet{scholte2026} have 10$^4$ K $<$ \te\oiii\ $<$ 1.4$\times$10$^4$ K, and the best-fit relation has a slope closer to unity. The LBT \yp\ Project nebulae at \te\oiii\ $<$ 1.4$\times$10$^4$ K also have \te\oiii\ $\approx$ \te\siii, which is similar to the \citet{garn1992} relation predictions in this temperature range. The \te\ scaling relations provided in Equations \ref{eq:ts3o3} and \ref{eq:to3s3} are only applicable over their calibration range, 1.1 $\lesssim$ \te\oiii/10$^4$ K $<$ 2.1, but argue for a shallow slope in \te\siii-\te\oiii.

Another striking feature is the low intrinsic dispersion about the best-fit relation. \te\siii\ has been observed to tightly correlate with \te\nii\ in high-metallicity nebulae \citep[e.g.,][]{berg2020}, but the scatter in \te\siii-\te\oiii\ measured from the same nebulae is typically $\sigma_{int}$ $\sim$ 830-1110 K \citep{roge2021,mend2023,scholte2026}. The intrinsic dispersion in Figure \ref{fig:ts3to3} is now comparable to the uncertainty on individual \te\oiii\ measurements, indicating that the intermediate and high-ionization zone temperatures are strongly related in the metal-poor ISM. A potential cause for the enhanced scatter in \te\ at high metallicities is contamination of the \oiii\ auroral line by \feii\ emission, although spectroscopic samples of extragalactic \hii\ regions have found that \feii\ contamination is infrequent \citep{roge2022,rick2024}.

An alternative explanation for the additional scatter in \te\siii-\te\oiii\ in metal-rich nebulae is a biased high-ionization zone temperature due to significant \te\ fluctuations within the ISM. In the presence of \te\ inhomogeneities, auroral line intensities are elevated and produce erroneously high \te\ when calculated under the assumption of uniform gas temperature \citep{peim1967}. However, observationally constraining the magnitude of \te\ fluctuations, which is parameterized by the value $t^2$, requires the detection of extremely faint metal recombination lines in the rest-optical \citep[e.g.,][]{este2004,este2009,este2020,tori2016,skil2020} or the \te-insensitive infrared fine-structure lines \citep{lama2022,chen2023}. The former are particularly challenging to detect in metal-poor nebulae, as the intensity of metal recombination lines scales with the abundance of the ion.
The \ion{O}{2}$\lambda$4639,4642 and \ion{O}{2}$\lambda$4649,4651 lines are detected in 8 and 3 nebulae, respectively, but are blended at the resolution of MODS. The \ion{O}{2}$\lambda$4639,4642 blend can be contaminated by other neighboring emission features \citep[see][]{este2004}, making these observations unsuitable for recombination-line abundances. Direct constraints on the temperature structure and $t^2$ within these metal-poor nebulae may be possible with followup observations on higher-resolution optical spectrographs or with infrared observations to detect the metal fine-structure lines.

Recently, \citet{mend2023N} found that the magnitude of $t^2$ is correlated with the difference between \te\nii\ and \te\oiii; under a simple geometric assumption of a \hii\ region, this trend may be related to larger \te\ fluctuations near the ionizing sources within the high-ionization zone. Additionally, they find that metal-poor, high-ionization nebulae similar to those in the LBT \yp\ Project are characterized by larger $t^2$, although only one extragalactic \hii\ region in their sample has \te\oiii\ $>$ 1.6$\times$10$^4$ K. The detection rate of the \nii\ auroral line is relatively low in the LBT \yp\ Project sample (see following subsection), so we cannot apply the \citet{mend2023N} relation to infer $t^2$ for the majority of the targets. If strong \te\ fluctuations are universally present in the nebulae, then one may expect \te\oiii\ to be biased high relative to the predictions of the \citet{garn1992} trend, as the photoionization model \te\ scaling relation is calibrated on the ion-weighted electron temperature and not the model line ratio temperatures. However, this does not mean the present sample is described by negligible $t^2$. Rather, temperature fluctuations may affect \te\oiii\ to a lesser extent than predicted by Eq 4 of \citet{mend2023N}, or that these variations also affect the measurement of \te\siii\ in the highly-ionized, high-\te\ ISM. To fully explore the shape of and scatter around the \te\siii-\te\oiii\ relation will require a larger sample of \hii\ regions that span a range of \te\ lower than those measured in the LBT \yp\ Project, but that is beyond the scope of the present analysis.

\subsection{\texorpdfstring{T$_e$\oiii}{TeO3} and Low-Ionization Zone Temperatures}\label{s:tlow}

\begin{figure*}[t]
   \centering
   \includegraphics[width=0.99\textwidth]{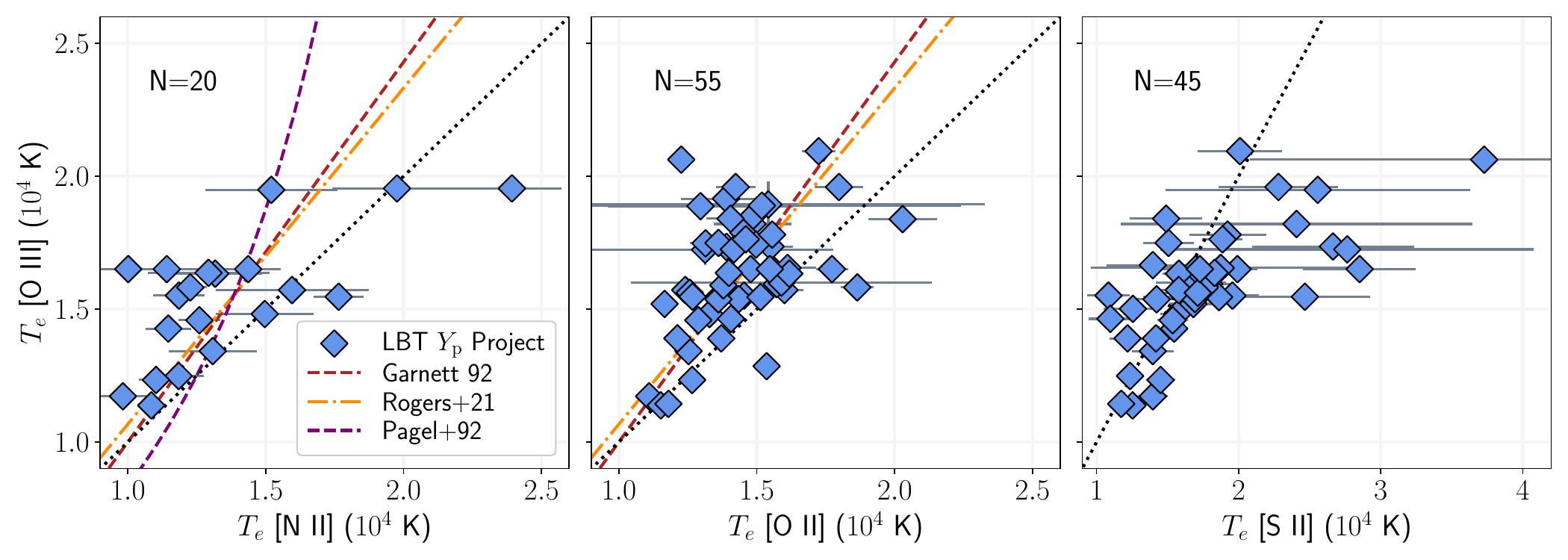}
   \caption{A comparison of \te\oiii\ in the LBT \yp\ Project sample to \te\ from the low-ionization emission lines of \nii\ (Left), \oii\ (Center), and \sii\ (Right). The \te\ scaling relations of \citet{garn1992}, \citet{page1992}, and \citet{roge2021} are plotted as red dashed, purple dot-dashed, and orange dashed lines, respectively. The black dotted line represents equivalent temperatures. The number of nebulae with simultaneous \te\ detections is provided in the upper left of each panel. The \te\oiii-\te\nii\ generally scatter around the literature \te\ scaling relations. \te\oiii-\te\oii\ and \te\oiii-\te\sii\ show substantial scatter to higher \te\oii\ and \te\sii\ relative to \te\oiii.}
   \label{fig:to3tlow}
\end{figure*}

The next most common auroral line detections are from \oii\ and \sii, with 55 and 45 significant detections, respectively. We also detect the \nii$\lambda$5755 auroral line in 20 targets, significantly less than the number of \oii\ auroral line detections owing to the low gas-phase N abundance and overall high ionization conditions in the ISM of low-metallicity star-forming galaxies. The \te\ trends of these three ions are plotted against \te\oiii\ in Figure \ref{fig:to3tlow}, where the color coding of the \te\ scaling relations are the same as used in Figure \ref{fig:ts3to3}. We have also included the \te\ scaling relation of \citet{page1992} as the purple dot-dashed line. We extrapolate these \te\ relations beyond their calibration range to assess the trends in the low-metallicity ISM.
The left panel reveals that the \te\oiii-\te\nii\ distribution generally follows the scaling relations of \citet{garn1992} and \citet{roge2021}; of the 20 nebulae with simultaneous \te\nii\ and \te\oiii, all but two have \te\oiii\ $\geq$ \te\nii. We do not attempt to calibrate a new \te\ scaling relation due to the paucity of simultaneous \te\nii\ and \te\oiii\ measurements.

The galaxies SHOC133 and WJ1205+4551 have a \te\nii\ that is $\sim$2000 K greater than the measured \te\oiii\ temperature. Both galaxies were observed by \citet{izot2021}, where SHOC133 is labeled therein as J0240$-$0828. In agreement with the previous observations, both SHOC133 and WJ1205+4551 show evidence of very high-ionization emission such as narrow \heii$\lambda$4686, [\ion{Ne}{5}]$\lambda$3426, and \fev$\lambda$4227. Our measurements of \te\oiii\ are in fairly good agreement with those reported in \citet{izot2021}, but there are no reported \te\nii\ in the prior observations of either galaxy. The MODS spectrum of SHOC133 also exhibits sharp line profiles with narrow super-Gaussian power P $\lesssim$ 1.7. Additionally, other studies have reported the detection of NIR coronal lines in WJ1205+4551 \citep{arav2024}; while the coronal lines of [\ion{Fe}{10}] are undetected in the LBT \yp\ Project MODS spectra of WJ1205+4551, there is a tentative detection of [\ion{Fe}{8}]$\lambda$6087 \citep[in agreement with][]{izot2021}.
It is interesting to note that the \te\siii\ in these galaxies are also greater than the measured \te\oiii, indicating an inverted \te\ structure relative to the predictions from photoionization models at low gas-phase metallicity.
Given the evidence from these spectral features and the inverted \te\ structure in the ISM, it is possible that other ionization sources significantly contribute to the CELs of SHOC133 and WJ1205+4551, despite their strong line ratios being indicative of an ionizing spectrum powered by stars.

With the exception of these two galaxies, the low-metallicity nebulae of the LBT \yp\ Project show \te\oiii\ $\gtrsim$ \te\nii\ when the faint \nii\ auroral line is detected. This is not the case for the temperature trends of \te\oiii\ vs.\ \te\oii\ and \te\sii, which are plotted in the center and right panels of Figure \ref{fig:to3tlow}, respectively. Prior studies have found large dispersion in empirical \te\oiii-\te\oii\ data \citep{kenn2003b,berg2020,yate2020,scholte2026}, and the standard deviation around the \citet{garn1992} relation observed in the LBT \yp\ Project \te\oii\ is $\sim$2600 K. 
While 17 nebulae have simultaneous \te\oiii\ and \te\oii\ that agree with the \citet{garn1992} scaling relation, \te\oii\ in 13 nebulae are greater than predicted by the scaling relation by more than 2$\sigma$.
This scatter could be influenced by several mechanisms; for example, the \oii\ auroral lines can redshift into the atmospheric A band at $\sim$7590-7710 \AA, but this would result in lower auroral line intensity and bias \te\oii\ low. We have already removed any nebulae that show significant absorption around the \oii\ auroral lines from Figure \ref{fig:to3tlow}, although increased night sky residuals in the NIR can complicate the auroral line detection and fitting for some of the nebulae.

The $^2$$P$ levels of the O$^+$ ion can be populated by recombination, resulting in increased \oii$\lambda$7320,30 emission and an erroneously high \te\oii\ if the $^2$$P$ levels are assumed to be populated purely by collisional excitation \citep{rubi1986,liu2001}. Equation 2 in \citet{liu2000} provides a means to account for the dielectronic recombination contribution of the \oii\ auroral line intensity relative to H$\beta$. We apply this equation to the subsample of nebulae with significant \oii\ auroral line detections and find that the average correction for dielectronic recombination is a $<$ 3\% reduction in the auroral line intensity \citep[consistent with other studies, see][]{kenn2003b}. This is because recombination scales with increasing O$^{2+}$ abundance in the ISM, but all LBT \yp\ Project targets have low total O abundance by selection.
The correction to the \oii\ auroral line intensity often produces statistically similar \te\oii, although the change in \te\ can be significant: the maximum predicted reduction in the \oii\ auroral line intensity is 7\% in the galaxy J1323$-$0132, which reduces \te\oii\ by $\sim$3000 K. Accounting for dielectronic recombination in the full sample, 11 nebulae remain 2$\sigma$ above the \citet{garn1992} \te\ scaling relation. 

While accounting for dielectronic recombination can lower \te\oii, it cannot explain the overall scatter in \te\oii\ observed in the full sample. Furthermore, no recombination contribution to the \sii$\lambda$4069,76 auroral lines has been reported, such that recombination cannot contribute to the scatter in \te\sii\ plotted in the right panel of Figure \ref{fig:to3tlow}. The \sii\ auroral-to-nebular line ratios in a handful of nebulae indicate \te\ in excess of 25 kK, far greater than \te\oiii\ or any other available temperature diagnostic. Given the large wavelength separation between the auroral and nebular lines, both \te\oii\ and \te\sii\ are sensitive to errors in the reddening correction. However, systematic uncertainties in $E(B-V)$ work in the opposite direction for the two temperatures, such that biases in $E(B-V)$ cannot explain the trend to high \te\sii\ and \te\oii.

In their work on the DESIRED sample, \citet{mend2023} discuss the possibility of \den\ inhomogeneities biasing the auroral-to-nebular line ratios of \oii\ and \sii. The critical densities of the $^2$$D$ levels of the O$^+$ and S$^+$ ions (on the order of 10$^3$ cm$^{-3}$ at \te$ = 1.5\times10^4$ K) are relatively small for optical CELs. If the ISM shows large \den\ variations, then collisional de-excitation of the nebular \oii\ and \sii\ lines could result in larger auroral-to-nebular line ratios and \te\ that are biased higher than predicted for the low-ionization zone. As described in \S\ref{s:dentrends} and plotted in Figure \ref{fig:ds2do2}, the auroral-to-nebular line ratios of \oii\ and \sii\ imply marginally higher average \den\ than the nebular line ratios, assuming a characteristic low-ionization zone \te. 
The scatter to high \te\oii\ and \te\sii\ in Figure \ref{fig:to3tlow} could be influenced by a combination of dielectronic recombination, reddening uncertainties, \den\ variations, or other factors not discussed here, such as shock excitation of the \sii\ lines or large \te\ variations in the low-ionization zone. Given the complexities associated with \te\oii\ and \te\sii, we recommend avoiding these temperatures for use in measuring chemical abundances in the ISM when other \te\ are available, in line with the recommendation of other recent works \citep{berg2020,mend2023}.

\section{Gas-Phase Oxygen Abundances}\label{s:loh}

\subsection{Ionic Abundance Calculations}\label{s:dirabun}

With \te\ and \den, we can now directly measure metal ionic abundances in the ISM of each LBT \yp\ Project target. In this work, we focus on the total abundance of O in the ISM, which is required to assess the relation between He/H and O/H and constrain \yp. In Paper VI, we will derive the chemical abundance patterns of many elements (from He to Fe) in a large sample of metal-poor nebulae.

To determine the ionic abundance of O$^+$ and O$^{2+}$ in the ISM, we select the \te\ measured in the MODS data that is most representative for the ionization zone of interest. For the high-ionization zone, we adopt the directly-measured \te\oiii, which is available for all targets in the sample.
For the low-ionization zone, we use \te\nii\ when available. \citet{kuri2021L} argue that a direct \te\oii\ is required to reliably measure the abundance of O$^+$/H$^+$ and its uncertainty, which can change significantly if a \te\ scaling relation is adopted. Indeed, \te\ scaling relations may not capture the simultaneous \te\ trends in the low-metallicity ISM, such as the \citet{page1992} relation in Figure \ref{fig:to3tlow}. However, \te\oii\ shows large dispersion around the \te\ scaling relations and can be biased by dielectronic recombination and density fluctuations in the ISM, as discussed in \S\ref{s:tlow}. These effects may introduce spurious trends in He/H vs.\ O/H and other relative abundance patterns, so we use the \te\oiii\ and the photoionization model scaling relation of \citet{garn1992} to infer the low-ionization zone temperature when \te\nii\ is unavailable. We propagate an additional 610 K into the inferred temperature uncertainty, which accounts for the scatter observed in empirical \te\nii\ and \te\oiii\ data \citep{roge2021} and mitigates the low uncertainties on O$^+$/H$^+$ discussed in \citet{kuri2021L}.
This may be an underestimate of the true dispersion in empirical \te\ in the low-metallicity nebulae, but we do not have the data to better constrain the dispersion or scaling relation for \te\nii-\te\oiii. Adopting \te\oii\ for the low-ionization zone \te\ can change the resulting 12+log(O/H) in individual objects, but does not alter the number of very low-metallicity galaxies that are critical for constraining \yp\ via a weighted average (see Paper IV).

Finally, we utilize the direct \te\siii\ for the intermediate-ionization zone, or apply Equation \ref{eq:ts3o3} to infer the intermediate-ionization zone \te\ when \siii$\lambda$6312 is not detected. Similar to the low-ionization zone, we consider $\sigma_{int}$ when inferring the intermediate-ionization zone temperature \citep[in the same manner as][]{roge2021}. The intermediate-ionization zone \te\ is not used for deriving O$^+$ and O$^{2+}$, but is necessary for future analysis of the S$^{2+}$, Ar$^{2+}$, and Cl$^{2+}$ abundances in the LBT \yp\ Project nebulae.

To measure ionic abundances relative to H$^+$, we apply the \textsc{PyNeb} \textsc{getIonAbundance} function using the emission line intensity ratio relative to H$\beta$, the measured or inferred ionization zone \te, and the adopted \den\ from \S\ref{s:tecalcs}. We calculate the uncertainty in the measured ionic abundance using the line intensity ratio uncertainty and the error on \te.
The ionic and total abundances of O calculated for the full sample are provided in Table \ref{t:abun}.

In general, it is not possible to detect emission from all relevant ionization states of a given element in the ISM. Correcting for unobserved ionization states, either through arguments of similar IP \citep{peim1969,crox2016} or through photoionization model predictions \citep{thua1995,izot2006,delg2014,amay2021}, represents a large systematic uncertainty present in most chemical abundance studies. Fortuitously, O is the only element with all relevant ionization states easily detectable within the MODS optical/NIR spectra. We assume that the total O abundance is O/H $=$ O$^{+}$/H$^+$ + O$^{2+}$/H$^+$, implying negligible contributions from neutral O or any higher ionization states. While \oi\ emission lines are clearly detected in the MODS spectra (see Figure \ref{fig:bptfigs}), this emission may not be cospatial with the ionized gas emitting the \hi\ recombination lines or where we have a direct \te\ measurement. Therefore, the inclusion of O$^0$/H$^+$ in the total O abundance is inappropriate for the present study. The ionization conditions of the LBT \yp\ Project sample are relatively high, with large O$_{32}$ and significant detections of narrow, nebular \heii$\lambda$4686, as well as other high-ionization emission lines such as \ariv$\lambda$47411,40 and \fev$\lambda$4227. It is natural to expect an amount of O$^{3+}$ in the ISM given these ionization conditions, but the detection of \ion{O}{4} emission lines is only possible with UV or MIR spectroscopy. In rare instances where \ion{O}{4} lines have been detected in extreme star-forming galaxies, the resulting ionic abundance of O$^{3+}$/H$^+$ in the ISM is only $\lesssim$2\% of the global O abundance \citep{berg2021,rick2025}. Such a contribution is less than the typical uncertainties on O/H measured in this sample, so we do not attempt to account for the O$^{3+}$ ions when reporting the total O/H. In addition to the emission line intensities and \te\oiii, the O/H abundance of each nebulae is provided for the final calibration of $Y$ vs.\ O/H to infer \yp\ (see Paper IV).

\subsection{O/H and Literature Comparison}\label{s:ohcomp}

\begin{figure*}[t]
   \centering
   \includegraphics[width=0.95\textwidth]{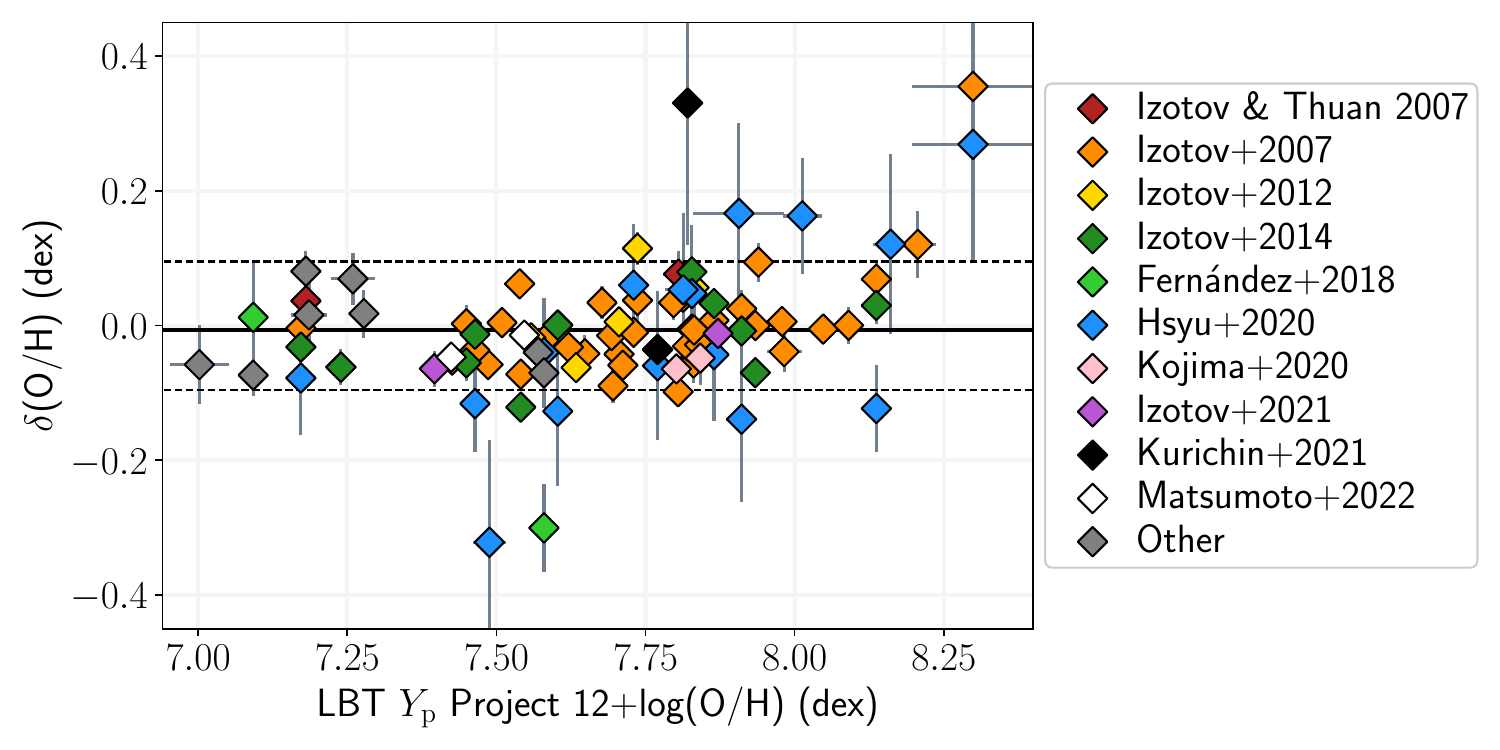}
   \caption{12+log(O/H) measured from the LBT/MODS spectra of the LBT \yp\ Project sample is plotted on the horizontal axis. The vertical axis plots $\delta$(O/H), or the difference between LBT \yp\ Project and literature 12+log(O/H) for the same galaxy (color-coded by source, see references in main text). The average offset in O/H is plotted as a solid black line, while the dashed lines represent the standard deviation in $\delta$(O/H) for all literature abundances. While the scatter in $\delta$(O/H) is larger than the typical error on O/H, and there are galaxies that show large variations in O/H, the measurement of 12+log(O/H) is generally consistent across observations, where the average $\delta$(O/H) is $-$0.006 dex.}
   \label{fig:dohoh}
\end{figure*}

The galaxies presented in this work were originally selected for their low gas-phase O/H, which is required to assess the relative abundance of He/H in gas of near primordial composition. A handful of moderately metal-enriched galaxies were included in the sample to explore the increase in He/H with O/H, a trend expected based on stellar nucleosynthesis and chemical evolution models \citep[e.g.,][]{well2025}. To measure He/H also requires many faint \hei\ recombination lines, some of which are fainter than the auroral lines necessary to directly measure \te\ and O/H. As such, fainter, potentially extremely metal-poor objects where \hei\ lines were unlikely to be robustly detected were de-prioritized in initial target selection. Given these selection criteria, it is unsurprising that we find the majority of the LBT \yp\ Project nebulae are metal poor, with 12+log(O/H) $<$ 20\% the solar O abundance. Specifically, the oxygen abundances measured from the LBT/MODS spectra span 12+log(O/H) $=$ [7.00,8.30] dex with median uncertainty of 4\%. The low uncertainty reflects the target selection and the quality of optical/NIR data accessible with LBT/MODS.

As a consequence of this methodology, we do not report many new metal-poor galaxies as part of this analysis. The exception is J0807+3414, which was selected based on its inferred metallicity of 12+log(O/H) $=$ 7.69$\pm$0.09 dex from \citet{sanc2016}. The direct-method metallicity measured in J0807+3414 is 12+log(O/H) $=$ 7.686$\pm$0.014 dex, in excellent agreement with the indirect estimate but with significantly lower uncertainty. For the remaining 61 galaxies, it is worthwhile to compare the direct O/H reported here to those from previous literature analyses that were considered when finalizing the target selection. In Figure \ref{fig:dohoh}, we plot the total O/H abundance measured as part of the LBT \yp\ Project on the horizontal axis. We define $\delta$(O/H) as the difference between the 12+log(O/H) measured in this work and the reported 12+log(O/H) of the same nebula from various literature sources. We color code each point based on the source of the previous O/H measurement, where the literature sources with multiple nebulae observed in the LBT \yp\ Project include: \citet[][red]{izot2007b}, \citet[][orange]{izot2007a}, \citet[][yellow]{izot2012}, \citet[][dark green]{izot2014}, \citet[][light green]{fern2018}, \citet[][blue]{hsyu2020}, \citet[][pink]{koji2020}, \citet[][purple]{izot2021}, \citet[][black]{kuri2021}, and  \citet[][white]{mats2022}. The gray points represent a compilation of literature sources with a single galaxy in common with the LBT \yp\ Project \citep[including][]{izot1999b,izot2006b,skil2013,hsyu2018,anni2019,aver2022,naka2022,wata2024}.

We find good consistency between the presented abundances and those in the literature, where the average $\delta$(O/H) is just $-$0.006 dex. However, it is notable that the standard deviation in $\delta$(O/H) is 0.095 dex, larger than the uncertainty on most 12+log(O/H) reported in the LBT \yp\ Project nebulae. Numerous factors can contribute to the difference in O/H, some of which are obvious in the largest outliers of Figure \ref{fig:dohoh}. For example, the galaxy UM420 has a metallicity of 12+log(O/H) $=$ 8.30$\pm$0.10 dex, the highest measured in the LBT \yp\ Project sample. The metallicity for UM420 reported in \citet{izot2007a} and \citet{hsyu2020} is $\sim$8.0 dex, hence the selection for observation as part of the LBT \yp\ Project. The $\sim$0.3 dex difference in O/H is related to the morphology of UM420. From integral field spectroscopy, \citet{jame2010} found that the two star-forming regions in UM420 have direct 12+log(O/H) $\approx$ 8.25$\pm$0.07 dex, in good agreement with the LBT \yp\ Project measurement. However, the integrated emission in UM420 results in 12+log(O/H) $=$ 8.03$^{+0.17}_{-0.13}$ dex, consistent with the measurements from \citet{izot2007a} and \citet{hsyu2020}. As illustrated in this example, differences in the aperture size and placement can incorporate different components of the ISM, resulting in differences in the integrated optical spectrum, measured physical conditions, and final 12+log(O/H) reported for individual galaxies.

Differences in methodology will also introduce offsets in the reported 12+log(O/H). Consider the galaxy HS0122+0743, which has 12+log(O/H) $=$ 7.580$\pm$0.015 dex as measured from the MODS spectrum. This metallicity is similar to that obtained from \citet{izot2007a}, \citet{hsyu2020}, and \citet{naka2022}, but is 0.33 dex lower than 12+log(O/H) reported in \citet{fern2018}. The \te\oiii\ reported in the latter is in good agreement with the \te\oiii\ measured from the MODS spectrum, but \citet{fern2018} adopted \te\siii, with the use of a \te\ scaling relation, for O$^{2+}$ ionic abundances on account of the lower temperature uncertainty. This produces a significantly different high-ionization zone \te\ that manifests as the large $\delta$(O/H) in this galaxy. Furthermore, direct measurements of \te\nii\ from the MODS spectra are possible in a portion of the LBT \yp\ Project sample (see Figure \ref{fig:to3tlow}), and these \te\nii\ are prioritized for the low-ionization zone \te\ and the measurement of O$^+$. Although O$^+$ is not the dominant ionization state of O, the high \te\nii\ measured in some of the LBT \yp\ Project nebulae results in lower total 12+log(O/H) than reported in the literature studies. An extreme example is the \te\ structure in  HS0837+4717: this galaxy has a direct \te\nii\ $=$ (1.98$\pm$0.24)$\times$10$^4$ K, which is over 3000 K higher than predicted from the \citet{garn1992} \te\ scaling relation and the direct \te\oiii. It is no surprise, then, that HS0837+4717 shows $\delta$(O/H) $\approx$ $-$0.12 dex when compared to the direct measurement presented in \citet{izot2014}. To a lesser extent, a change to the adopted density could also contribute to a change in the O$^+$ ionic abundance and $\delta$(O/H), although most of the LBT \yp\ Project targets are close to the low-density limit (Figure \ref{fig:ne-ne}).

Numerous other factors could produce variations in 12+log(O/H) reported here and in the literature. Some of these include differences in the adopted atomic data \citep{berg2015,mori2020}, the use of the \oii\ auroral lines for O$^+$ abundances \citep{rodr2020}, and corrections due to the contamination of \oiii$\lambda$4363 from \feii\ emission \citep[][although this is generally a small correction, see \S\ref{s:linefit}]{curt2017,roge2022}. In summary, although individual measurements of 12+log(O/H) in a given galaxy can vary significantly, the metallicity measured from the LBT/MODS spectroscopy and reported in the literature are generally consistent and indicate that the sample is metal-poor in nature. The utility of the LBT \yp\ Project sample is in its homogeneity, where the 62 galaxies have been observed with the same instruments and analyzed self-consistently to produce a dataset capable of robustly constraining the primordial He abundance.

\section{Conclusions}\label{s:conclusions}

In this paper, we describe the LBT/MODS observations of the galaxies observed as part of the LBT \yp\ Project. These galaxies have been selected for observation based on their low gas-phase metallicities and large \ion{H}{1} EWs. The large, homogeneous sample provides an opportunity to measure the physical conditions in the low-metallicity ISM of local star-forming galaxies. Our key findings are:
\begin{enumerate}
    \item We find that the emission lines in the average MODS spectrum are not described by Gaussian profiles, a result of the telescope optics. We implement a new emission line fitting methodology, employing super-Gaussian emission line profiles that better capture the position, amplitude, and shape of the emission lines measured in MODS (Equations \ref{eq:sg} and \ref{eq:sg_sig}). Furthermore, we fit two broad Gaussian profiles to the H$\beta$, \oiii$\lambda$4959,5007, and H$\alpha$ emission lines in a majority of the LBT \yp\ Project galaxies (Figure \ref{fig:sg_comp}). These more complex line profiles are crucial for faint emission lines neighboring bright features (e.g., \hei\ $\lambda$5015 and \nii$\lambda$6548,84), where the flux of the faint lines may be biased too high if the broad wings are included in the narrow line flux. Finally, we preserve the native resolution of MODS, allowing for a simultaneous measurement of \oii$\lambda$3726,29 and \den\oii. 
    \item The ionization conditions, as traced by various emission line ratios, in the LBT \yp\ Project nebulae are very high. The positions of the galaxies in the traditional BPT diagrams are consistent with very low gas-phase O/H (Figure \ref{fig:bptfigs}), while the relative metal line ratios are similar to those measured in high-\emph{z} extreme emission line galaxies (Figure \ref{fig:extra_bptfigs}). There are 50 galaxies with significant, narrow, nebular \heii$\lambda$4686 emission, ranging from 0.2 to 3.4\% the intensity of H$\beta$. A subset of galaxies also show the very high-ionization emission from \fev$\lambda$4227. Such highly ionized gas is challenging to produce from stars alone, and may hint at more complicated physics in the ISM.
    \item We derive the physical conditions, \te\ and \den, in the ISM of the LBT \yp\ Project targets. This is an important step for the measurement of \yp: \te\oiii\ acts as a weak prior on \te(\ion{He}{1}), and both \te\ and \den\ are required to measure the gas-phase abundance of O$^+$ and O$^{2+}$. The simultaneous \den\sii\ and \den\oii\ generally agree, although the marginally lower average \den\sii\ could be related to DIG contamination of the \sii\ strong lines. Both \den\sii\ and \den\oii\ are significantly lower than the densities measured from the \cliii\ and \ariv\ emission lines (Figure \ref{fig:ne-ne}). This offset persists when using the \oii\ and \sii\ auroral-to-nebular line ratios as density tracers, which do predict generally larger \den\oii\ than the nebular lines alone (Figure \ref{fig:ds2do2}). However, the use of the densities from higher ionization ions such as Ar$^{3+}$ does not significantly affect \te\oiii\ or O$^{2+}$ abundances. 
    \item
    The high-quality MODS spectra resolve the \te\ structure in the multi-phase ISM, and we find a strong correlation between \te\siii\ and \te\oiii\ from 60 nebulae (Figure \ref{fig:ts3to3}). The resulting \te\ scaling relations (Equations \ref{eq:ts3o3} and \ref{eq:to3s3}) can be used to infer the temperature of the intermediate- and high-ionization zones when a direct \te\ is unavailable in the metal-poor, high-ionization ISM. The small scatter about this relation is in contrast to the \te\siii\ and \te\oiii\ trends of more metal-rich \hii\ regions. Combinations with larger \te\ samples in metal-rich nebulae will assess the shape of these relations and the drivers of scatter in \te\siii\ and \te\oiii.
    \item \te\nii\ is measured in fewer galaxies, a result of the low metallicities and high ionization conditions in the present sample. Nevertheless, we find that the direct \te\oiii-\te\nii\ tend to scatter around the \te\ scaling relations of \citet{garn1992} and \citet{roge2021}. However, there are a handful of LBT \yp\ Project targets with very high \te\nii $\gtrsim$ 2$\times$10$^4$ K, which are not predicted by empirical or photoionization model \te\ scaling relations. Furthermore, multiple galaxies show elevated \te\oii\ and \te\sii\ relative to \te\oiii\ (Figure \ref{fig:to3tlow}), a trend that could be related to the density sensitivity of the \oii\ and \sii\ nebular lines or other mechanisms of producing the auroral lines of these ionic species. Given these systematic uncertainties and the large observed scatter in \te, we caution against the use of \te\oii\ and \te\sii\ for ionic abundance measurements when other \te\ are available.
    \item With the \te, \den, and the collisionally excited metal lines available in the optical, we measure the total O abundance in the ISM of the LBT \yp\ Project galaxies. Given the high-fidelity constraints on \te\oiii, the average uncertainty on the O/H is just 4\%. We compare the direct O/H in individual objects to measurements from various literature sources, finding an average offset of $\delta$(O/H) $=$ $-$0.006 dex with a standard deviation of 0.095 dex (Figure \ref{fig:dohoh}). While 12+log(O/H) can vary significantly for some objects (as discussed in \S\ref{s:ohcomp}), the O/H measurements are, on average, consistent across observations.
\end{enumerate}

The data presented in this manuscript provide the optical \hi\ and \hei\ emission lines fluxes and \te\oiii\ required to determine the ISM He/H abundance, measure the He/H-O/H relation, and constrain \yp, as will be detailed in Paper IV of this series. The LBT/MODS spectra of the LBT \yp\ Project galaxies reveal the presence of emission lines from many other elements in the ISM. In Paper VI of this series, we will use the physical conditions derived in this analysis to examine the chemical abundance patterns, from He to Fe, in the low-metallicity ISM. Such a sample will provide a necessary comparison sample for low-metallicity galaxies being observed with JWST. Furthermore, we will determine if the peculiar emission line ratios or physical conditions, such as \te\nii\ $>$ \te\oiii, are indicators of anomalous gas-phase abundance patterns.

\begin{acknowledgements}
We thank the referee for their useful feedback and recommendations, which we believe have strengthened the manuscript.
This work was supported by funds provided by NSF Collaborative Research Grants AST-2205817 to RWP, AST-2205864 to EDS, and AST-2205958 to EA.
Our team workshop at OSU in July 2024 was sponsored in part by OSU's Center for Cosmology and AstroParticle Physics (CCAPP).

This work is based on observations made with the Large Binocular Telescope. The LBT is an international collaboration among institutions in the United States, Italy and Germany. LBT Corporation Members are: The University of Arizona on behalf of the Arizona Board of Regents; Istituto Nazionale di Astrofisica, Italy; LBT Beteiligungsgesellschaft, Germany, representing the Max-Planck Society, The Leibniz Institute for Astrophysics Potsdam, and Heidelberg University; The Ohio State University, and The Research Corporation, on behalf of The University of Notre Dame, University of Minnesota and University of Virginia. Observations have benefited from the use of ALTA Center (alta.arcetri.inaf.it) forecasts performed with the Astro-Meso-Nh model. Initialization data of the ALTA automatic forecast system come from the General Circulation Model (HRES) of the European Centre for Medium Range Weather Forecasts.

This research used the facilities of the Italian Center for Astronomical Archive (IA2) operated by INAF at the Astronomical Observatory of Trieste.

EDS, EA, DAB, NSJR, and JHM would like to acknowledge and thank Stanley Hubbard for his generous gift to
the University of Minnesota that allowed the University to become a member of the LBT collaboration.

\end{acknowledgements}

\facilities{LBT (MODS), LBT (LUCI)}
\software{
\texttt{astropy} \citep{astr2013, astr2018, astr2022},
\texttt{jupyter} \citep{kluy2016},
\texttt{modsIDL} \citep{crox2019},
\texttt{modsCCDRed} \citep{pogg2019},
\texttt{PypeIt} \citep{pypeit:joss_pub, pypeit:zenodo}
\texttt{PyNeb} \citep{luri2012,luri2015L},
\texttt{numpy} \citep{harr2020},
\texttt{scipy} \citep{2020SciPy}}

\appendix
\vspace{-2ex}
\section{Measured Properties}\label{app:measurements}

In this appendix, we provide tables with the measured properties in the LBT \yp\ Project sample. H and He recombination line flux ratios necessary for He abundances are included in Table \ref{t:hiinten}, reddening-corrected line intensity measurements in the MODS spectra are reported in Table \ref{t:allinten}, and the direct \den, \te, ionic and total abundances of O are presented in Table \ref{t:abun}.


\startlongtable
\begin{deluxetable*}{lccccccc}  
\tablecaption{LBT \yp\ Project H and He Line Flux Ratios and EWs \label{t:hiinten}}
\tablewidth{\columnwidth}
\tabletypesize{\footnotesize}
\tablehead{
   \colhead{Line} & 
   \colhead{AGC198691} & 
   \colhead{DDO68} & 
   \colhead{HS0029+1748} &
   \colhead{HS0122+0743} &
   \colhead{HS0134+3415} &
   \colhead{HS0811+4913} &
   \colhead{HS0837+4717}}
\startdata
{H12 $\lambda$3750  } & 0.0108$\pm$0.0071 & 0.0246$\pm$0.0035 & 0.0203$\pm$0.0016 & 0.0236$\pm$0.0011 & 0.0247$\pm$0.0005 & 0.0227$\pm$0.0013 & 0.0246$\pm$0.0018 \\
 EW  & 0.56 & 1.50 & 2.65 & 4.05 & 5.70 & 6.52 & 3.17 \\
\hline
{H11 $\lambda$3771  } & 0.0167$\pm$0.0073 & 0.0293$\pm$0.0034 & 0.0259$\pm$0.0018 & 0.0301$\pm$0.0017 & 0.0320$\pm$0.0007 & 0.0291$\pm$0.0011 & 0.0304$\pm$0.0015 \\
 EW  & 0.85 & 1.81 & 3.38 & 5.14 & 7.43 & 8.48 & 3.92 \\
\hline
{H10 $\lambda$3797  } & 0.0228$\pm$0.0068 & 0.0423$\pm$0.0036 & 0.0355$\pm$0.0018 & 0.0405$\pm$0.0017 & 0.0436$\pm$0.0008 & 0.0409$\pm$0.0010 & 0.0418$\pm$0.0014 \\
 EW  & 1.15 & 2.64 & 4.60 & 6.89 & 10.17 & 12.14 & 5.41 \\
\hline
{H9 $\lambda$3835  } & 0.0434$\pm$0.0062 & 0.0580$\pm$0.0032 & 0.0495$\pm$0.0016 & 0.0554$\pm$0.0013 & 0.0617$\pm$0.0010 & 0.0564$\pm$0.0011 & 0.0573$\pm$0.0014 \\
 EW  & 2.16 & 3.69 & 6.38 & 9.35 & 14.54 & 17.22 & 7.41 \\
\hline
{\ion{He}{1}+H8 $\lambda$3889  } & 0.1516$\pm$0.0049 & 0.1950$\pm$0.0046 & 0.1542$\pm$0.0024 & 0.1648$\pm$0.0023 & 0.1678$\pm$0.0024 & 0.1600$\pm$0.0024 & 0.1521$\pm$0.0025 \\
 EW  & 7.60 & 12.38 & 20.34 & 28.85 & 40.46 & 50.37 & 20.49 \\
\hline
{\ion{He}{1} $\lambda$4026  } & 0.0134$\pm$0.0043 & 0.0107$\pm$0.0025 & 0.0145$\pm$0.0009 & 0.0148$\pm$0.0007 & 0.0158$\pm$0.0004 & 0.0145$\pm$0.0004 & 0.0150$\pm$0.0008 \\
 EW  & 0.76 & 0.75 & 1.93 & 2.74 & 3.96 & 4.84 & 2.17 \\
\hline
{H$\delta$ $\lambda$4101  } & 0.2232$\pm$0.0059 & 0.2556$\pm$0.0048 & 0.2146$\pm$0.0033 & 0.2259$\pm$0.0034 & 0.2329$\pm$0.0036 & 0.2223$\pm$0.0035 & 0.2309$\pm$0.0035 \\
 EW  & 12.35 & 18.43 & 29.22 & 43.15 & 61.14 & 75.35 & 34.56 \\
\hline
{H$\gamma$ $\lambda$4340  } & 0.4485$\pm$0.0083 & 0.4669$\pm$0.0077 & 0.4161$\pm$0.0062 & 0.4308$\pm$0.0064 & 0.4401$\pm$0.0068 & 0.4149$\pm$0.0064 & 0.4308$\pm$0.0066 \\
 EW  & 28.34 & 40.89 & 63.62 & 92.14 & 129.91 & 159.01 & 74.17 \\
\hline
{\ion{He}{1} $\lambda$4388  } & \nodata & 0.0025$\pm$0.0022 & 0.0040$\pm$0.0006 & 0.0041$\pm$0.0009 & 0.0042$\pm$0.0003 & 0.0040$\pm$0.0004 & 0.0038$\pm$0.0013 \\
 EW  & \nodata & 0.23 & 0.61 & 0.89 & 1.26 & 1.55 & 0.67 \\
\hline
{\ion{He}{1} $\lambda$4471  } & 0.0307$\pm$0.0041 & 0.0324$\pm$0.0027 & 0.0350$\pm$0.0008 & 0.0348$\pm$0.0010 & 0.0367$\pm$0.0006 & 0.0347$\pm$0.0006 & 0.0388$\pm$0.0014 \\
 EW  & 2.04 & 3.11 & 5.53 & 7.88 & 11.49 & 14.20 & 7.16 \\
\hline
{\ion{He}{2} $\lambda$4686  } & \nodata & 0.0274$\pm$0.0019 & 0.0087$\pm$0.0005 & 0.0097$\pm$0.0006 & 0.0058$\pm$0.0003 & 0.0019$\pm$0.0004 & 0.0190$\pm$0.0009 \\
 EW  & \nodata & 3.04 & 0.24 & 2.39 & 1.97 & 0.83 & 3.85 \\
\hline
{\ion{He}{1} $\lambda$4922  } & 0.0060$\pm$0.0037 & 0.0082$\pm$0.0020 & 0.0103$\pm$0.0006 & 0.0103$\pm$0.0006 & 0.0100$\pm$0.0004 & 0.0095$\pm$0.0003 & 0.0121$\pm$0.0019 \\
 EW  & 0.47 & 1.05 & 1.88 & 2.82 & 3.73 & 4.62 & 2.74 \\
\hline
{\ion{He}{1} $\lambda$5016  } & 0.0216$\pm$0.0037 & 0.0192$\pm$0.0021 & 0.0203$\pm$0.0007 & 0.0204$\pm$0.0007 & 0.0173$\pm$0.0005 & 0.0191$\pm$0.0004 & 0.0255$\pm$0.0019 \\
 EW  & 1.78 & 2.60 & 3.80 & 5.82 & 6.76 & 9.69 & 6.03 \\
\hline
{\ion{He}{1} $\lambda$5876  } & 0.1106$\pm$0.0042 & 0.0952$\pm$0.0022 & 0.1181$\pm$0.0024 & 0.1179$\pm$0.0019 & 0.1106$\pm$0.0018 & 0.1336$\pm$0.0021 & 0.1423$\pm$0.0022 \\
 EW  & 12.48 & 22.37 & 29.79 & 45.85 & 59.01 & 88.90 & 45.98 \\
\hline
{H$\alpha$ $\lambda$6563  } & 3.0802$\pm$0.0477 & 2.9633$\pm$0.0471 & 3.3047$\pm$0.0491 & 3.3497$\pm$0.0496 & 3.1314$\pm$0.0492 & 3.7694$\pm$0.0590 & 3.8129$\pm$0.0581 \\
 EW  & 442.89 & 917.92 & 956.89 & 1492.35 & 1940.75 & 2841.61 & 1433.58 \\
\hline
{\ion{He}{1} $\lambda$6678  } & 0.0344$\pm$0.0022 & 0.0272$\pm$0.0007 & 0.0357$\pm$0.0006 & 0.0355$\pm$0.0007 & 0.0324$\pm$0.0005 & 0.0416$\pm$0.0007 & 0.0380$\pm$0.0007 \\
 EW  & 5.18 & 8.78 & 10.45 & 16.17 & 20.32 & 31.29 & 14.60 \\
\hline
{\ion{He}{1} $\lambda$7065  } & 0.0257$\pm$0.0046 & 0.0238$\pm$0.0013 & 0.0295$\pm$0.0005 & 0.0430$\pm$0.0008 & 0.0329$\pm$0.0005 & 0.0431$\pm$0.0007 & 0.0879$\pm$0.0015 \\
 EW  & 4.08 & 8.82 & 9.54 & 22.44 & 23.55 & 36.26 & 39.47 \\
\hline
{\ion{He}{1} $\lambda$7281  } & 0.0036$\pm$0.0046 & 0.0055$\pm$0.0011 & 0.0063$\pm$0.0007 & 0.0085$\pm$0.0007 & 0.0069$\pm$0.0002 & 0.0089$\pm$0.0003 & 0.0074$\pm$0.0067 \\
 EW  & 0.62 & 2.22 & 2.13 & 4.63 & 5.13 & 7.92 & 3.38 \\
\hline
{P15 $\lambda$8545} & \nodata & 0.0058$\pm$0.0010 & 0.0054$\pm$0.0006 & 0.0055$\pm$0.0007 & 0.0064$\pm$0.0003 & 0.0092$\pm$0.0003 & 0.0070$\pm$0.0016 \\
 EW  & \nodata & 4.04 & 2.70 & 4.62 & 7.99 & 17.62 & 5.31 \\
\hline
{P14 $\lambda$8598} & \nodata & 0.0081$\pm$0.0011 & 0.0078$\pm$0.0008 & 0.0080$\pm$0.0007 & 0.0082$\pm$0.0002 & 0.0112$\pm$0.0004 & 0.0109$\pm$0.0013 \\
 EW  & \nodata & 5.91 & 3.92 & 6.89 & 10.50 & 21.79 & 8.33 \\
\hline
{P13 $\lambda$8665} & 0.0268$\pm$0.0111 & 0.0094$\pm$0.0012 & 0.0098$\pm$0.0010 & 0.0109$\pm$0.0006 & 0.0097$\pm$0.0003 & 0.0140$\pm$0.0004 & 0.0122$\pm$0.0008 \\
 EW  & 6.32 & 7.17 & 4.95 & 9.55 & 12.63 & 27.98 & 9.53 \\
\hline
{P12 $\lambda$8750} & \nodata & 0.0096$\pm$0.0045 & 0.0138$\pm$0.0010 & 0.0136$\pm$0.0008 & 0.0131$\pm$0.0004 & 0.0183$\pm$0.0025 & 0.0143$\pm$0.0006 \\
 EW  & \nodata & 7.76 & 7.06 & 12.27 & 17.52 & 37.73 & 11.49 \\
\hline
{P11 $\lambda$8863} & 0.0158$\pm$0.0104 & 0.0152$\pm$0.0057 & 0.0171$\pm$0.0008 & 0.0175$\pm$0.0010 & 0.0168$\pm$0.0005 & 0.0232$\pm$0.0032 & 0.0201$\pm$0.0007 \\
 EW  & 4.19 & 13.36 & 8.93 & 16.31 & 23.15 & 49.84 & 16.62 \\
\hline
{P10 $\lambda$9015} & 0.0233$\pm$0.0039 & 0.0199$\pm$0.0019 & 0.0217$\pm$0.0006 & 0.0272$\pm$0.0006 & 0.0219$\pm$0.0006 & 0.0298$\pm$0.0008 & 0.0248$\pm$0.0012 \\
 EW  & 6.65 & 16.38 & 11.27 & 24.93 & 29.55 & 57.23 & 18.07 \\
\hline
{P9 $\lambda$9229} & 0.0215$\pm$0.0041 & 0.0284$\pm$0.0015 & 0.0294$\pm$0.0023 & 0.0262$\pm$0.0039 & 0.0308$\pm$0.0013 & 0.0423$\pm$0.0013 & 0.0372$\pm$0.0026 \\
 EW  & 6.51 & 25.77 & 16.65 & 26.88 & 46.43 & 92.28 & 30.14 \\
\hline
{P8 $\lambda$9546} & 0.0430$\pm$0.0186 & 0.0368$\pm$0.0057 & 0.0539$\pm$0.0019 & 0.0527$\pm$0.0018 & 0.0419$\pm$0.0020 & 0.0623$\pm$0.0025 & 0.0458$\pm$0.0025 \\
 EW  & 12.40 & 41.24 & 32.48 & 54.63 & 67.41 & 143.15 & 39.14 \\
\hline
{\emph{z}} & 0.00171 & 0.00168 & 0.00707 & 0.00976 & 0.01949 & 0.00185 & 0.04197 \\
{F$_{H\beta}$} &  9.5$\pm$ 0.1 & 25.3$\pm$ 0.3 & 141.6$\pm$ 1.5 & 114.0$\pm$ 1.2 & 202.4$\pm$ 2.2 & 142.3$\pm$ 1.6 & 159.9$\pm$ 1.7 \\
{EW$_{H\beta}$} & 76.88 & 122.19 & 179.42 & 266.04 & 365.41 & 472.55 & 219.50 \\
{$E(B-V)$} & 0.10$\pm$0.01 & 0.06$\pm$0.01 & 0.15$\pm$0.01 & 0.19$\pm$0.01 & 0.12$\pm$0.01 & 0.27$\pm$0.01 & 0.26$\pm$0.01 \\
{$a_H$} & 1.4$_{-0.4}^{+0.4}$ & 0.0$_{-0.0}^{+0.4}$ & 3.5$_{-0.5}^{+0.5}$ & 0.3$_{-0.3}^{+0.8}$ & 2.8$_{-1.2}^{+1.2}$ & 0.0$_{-0.0}^{+1.4}$ & 0.0$_{-0.0}^{+0.6}$ \\
\enddata
\tablecomments{Emission line fluxes, uncorrected for reddening, required to measure He/H. Each row corresponds to the observed flux of a \hi, \hei, or \heii\ recombination line relative to the observed flux of H$\beta$ (top) and that line's EW (in \AA, bottom). The last five rows provide the spectroscopic \emph{z}, observed flux of H$\beta$ (in 10$^{-16}$ erg/s/cm$^{2}$), the EW of H$\beta$ (in \AA), $E(B-V)$ and Balmer absorption EW, $a_H$, inferred from the methods of \S\ref{s:redcorr}. Table truncated at seven targets, the complete version is available in machine readable format.}
\end{deluxetable*}

\startlongtable
\begin{deluxetable*}{lccccccc}  
\tablecaption{LBT \yp\ Project Line Intensities \label{t:allinten}}
\tablewidth{\columnwidth}
\tabletypesize{\footnotesize}
\tablehead{
   \colhead{Line} & 
   \colhead{AGC198691} & 
   \colhead{DDO68} & 
   \colhead{HS0029+1748} &
   \colhead{HS0122+0743} &
   \colhead{HS0134+3415} &
   \colhead{HS0811+4913} &
   \colhead{HS0837+4717}}
\startdata
{[\ion{Ne}{3}]$\lambda$3342  } & \nodata & 0.0028$\pm$0.0075 & \nodata & 0.0024$\pm$0.0032 & 0.0021$\pm$0.0011 & 0.0019$\pm$0.0020 & 0.0044$\pm$0.0031 \\
{[\ion{Ne}{5}]$\lambda$3346  } & \nodata & 0.0061$\pm$0.0075 & 0.0003$\pm$0.0027 & 0.0025$\pm$0.0032 & \nodata & 0.0030$\pm$0.0020 & 0.0011$\pm$0.0031 \\
{[\ion{Ne}{5}]$\lambda$3426  } & 0.0022$\pm$0.0228 & \nodata & 0.0006$\pm$0.0024 & 0.0016$\pm$0.0027 & \nodata & \nodata & 0.0063$\pm$0.0028 \\
{H14 $\lambda$3722  } & 0.0194$\pm$0.0088 & 0.0228$\pm$0.0049 & 0.0335$\pm$0.0015 & 0.0275$\pm$0.0012 & 0.0296$\pm$0.0008 & 0.0369$\pm$0.0020 & 0.0373$\pm$0.0029 \\
{[\ion{O}{2}]$\lambda$3726  } & 0.1693$\pm$0.0095 & 0.1636$\pm$0.0060 & 0.3313$\pm$0.0065 & 0.2914$\pm$0.0057 & 0.1704$\pm$0.0035 & 0.4519$\pm$0.0096 & 0.2469$\pm$0.0056 \\
{[\ion{O}{2}]$\lambda$3729  } & 0.2473$\pm$0.0103 & 0.2416$\pm$0.0071 & 0.4805$\pm$0.0093 & 0.4147$\pm$0.0080 & 0.2329$\pm$0.0048 & 0.5939$\pm$0.0125 & 0.2653$\pm$0.0059 \\
{H13 $\lambda$3734  } & 0.0197$\pm$0.0088 & 0.0237$\pm$0.0049 & 0.0218$\pm$0.0015 & 0.0277$\pm$0.0012 & 0.0224$\pm$0.0007 & 0.0283$\pm$0.0019 & 0.0292$\pm$0.0029 \\
{H12 $\lambda$3750  } & 0.0117$\pm$0.0078 & 0.0262$\pm$0.0038 & 0.0233$\pm$0.0019 & 0.0289$\pm$0.0014 & 0.0277$\pm$0.0007 & 0.0301$\pm$0.0017 & 0.0324$\pm$0.0024 \\
{H11 $\lambda$3771  } & 0.0181$\pm$0.0079 & 0.0312$\pm$0.0036 & 0.0297$\pm$0.0021 & 0.0367$\pm$0.0021 & 0.0358$\pm$0.0009 & 0.0384$\pm$0.0015 & 0.0398$\pm$0.0021 \\
{H10 $\lambda$3797  } & 0.0247$\pm$0.0074 & 0.0450$\pm$0.0039 & 0.0405$\pm$0.0021 & 0.0493$\pm$0.0022 & 0.0487$\pm$0.0011 & 0.0537$\pm$0.0015 & 0.0546$\pm$0.0020 \\
{\ion{He}{1} $\lambda$3820  } & 0.0074$\pm$0.0067 & 0.0078$\pm$0.0033 & 0.0079$\pm$0.0016 & 0.0073$\pm$0.0011 & 0.0077$\pm$0.0004 & 0.0090$\pm$0.0007 & 0.0113$\pm$0.0015 \\
{H9 $\lambda$3835  } & 0.0468$\pm$0.0067 & 0.0615$\pm$0.0035 & 0.0564$\pm$0.0019 & 0.0670$\pm$0.0017 & 0.0687$\pm$0.0014 & 0.0736$\pm$0.0016 & 0.0742$\pm$0.0021 \\
{[\ion{Ne}{3}]$\lambda$3869  } & 0.1044$\pm$0.0040 & 0.1629$\pm$0.0043 & 0.4106$\pm$0.0076 & 0.3748$\pm$0.0069 & 0.5894$\pm$0.0115 & 0.6116$\pm$0.0120 & 0.5422$\pm$0.0103 \\
{\ion{He}{1}+H8 $\lambda$3889  } & 0.1630$\pm$0.0047 & 0.2063$\pm$0.0050 & 0.1745$\pm$0.0033 & 0.1977$\pm$0.0037 & 0.1858$\pm$0.0036 & 0.2064$\pm$0.0040 & 0.1948$\pm$0.0039 \\
{\ion{He}{1} $\lambda$3965  } & 0.0031$\pm$0.0053 & 0.0034$\pm$0.0025 & 0.0042$\pm$0.0011 & 0.0050$\pm$0.0009 & 0.0059$\pm$0.0004 & 0.0075$\pm$0.0005 & 0.0084$\pm$0.0008 \\
{[\ion{Ne}{3}]$\lambda$3967  } & 0.0228$\pm$0.0054 & 0.0529$\pm$0.0027 & 0.1352$\pm$0.0026 & 0.1115$\pm$0.0022 & 0.1733$\pm$0.0033 & 0.1865$\pm$0.0036 & 0.1597$\pm$0.0030 \\
{H7 $\lambda$3970  } & 0.1348$\pm$0.0060 & 0.1714$\pm$0.0042 & 0.1361$\pm$0.0026 & 0.1598$\pm$0.0030 & 0.1618$\pm$0.0031 & 0.1611$\pm$0.0031 & 0.1768$\pm$0.0033 \\
{\ion{He}{1} $\lambda$4026  } & 0.0142$\pm$0.0045 & 0.0113$\pm$0.0026 & 0.0161$\pm$0.0010 & 0.0173$\pm$0.0009 & 0.0173$\pm$0.0005 & 0.0181$\pm$0.0006 & 0.0186$\pm$0.0011 \\
{[\ion{S}{2}]$\lambda$4069  } & 0.0046$\pm$0.0051 & 0.0042$\pm$0.0028 & 0.0096$\pm$0.0691 & 0.0078$\pm$0.0008 & 0.0065$\pm$0.0004 & 0.0090$\pm$0.0005 & 0.0108$\pm$0.0010 \\
{[\ion{S}{2}]$\lambda$4076  } & 0.0018$\pm$0.0051 & 0.0023$\pm$0.0028 & 0.0015$\pm$0.0008 & 0.0031$\pm$0.0007 & 0.0015$\pm$0.0004 & 0.0031$\pm$0.0005 & 0.0029$\pm$0.0009 \\
{H$\delta$ $\lambda$4101  } & 0.2613$\pm$0.0101 & 0.2673$\pm$0.0079 & 0.2624$\pm$0.0061 & 0.2631$\pm$0.0055 & 0.2630$\pm$0.0065 & 0.2727$\pm$0.0067 & 0.2815$\pm$0.0067 \\
{\ion{He}{1} $\lambda$4121  } & 0.0027$\pm$0.0051 & \nodata & 0.0005$\pm$0.0007 & 0.0023$\pm$0.0008 & 0.0015$\pm$0.0003 & 0.0020$\pm$0.0004 & 0.0033$\pm$0.0008 \\
{\ion{He}{1} $\lambda$4144  } & 0.0020$\pm$0.0047 & \nodata & 0.0017$\pm$0.0006 & 0.0018$\pm$0.0005 & 0.0028$\pm$0.0004 & 0.0023$\pm$0.0004 & 0.0048$\pm$0.0011 \\
{\ion{He}{1} $\lambda$4169  } & 0.0014$\pm$0.0040 & \nodata & 0.0001$\pm$0.0005 & 0.0003$\pm$0.0005 & 0.0003$\pm$0.0004 & 0.0005$\pm$0.0004 & \nodata \\
{[\ion{Fe}{5}]$\lambda$4227  } & 0.0058$\pm$0.0040 & 0.0017$\pm$0.0032 & \nodata & 0.0007$\pm$0.0005 & 0.0010$\pm$0.0004 & 0.0001$\pm$0.0004 & 0.0026$\pm$0.0012 \\
{\ion{C}{2} $\lambda$4267  } & \nodata & 0.0010$\pm$0.0021 & 0.0009$\pm$0.0007 & \nodata & 0.0004$\pm$0.0003 & \nodata & 0.0003$\pm$0.0010 \\
{[\ion{Fe}{2}]$\lambda$4287  } & \nodata & 0.0020$\pm$0.0021 & 0.0007$\pm$0.0007 & 0.0032$\pm$0.0008 & 0.0004$\pm$0.0003 & 0.0001$\pm$0.0004 & 0.0023$\pm$0.0010 \\
{H$\gamma$ $\lambda$4340  } & 0.4854$\pm$0.0115 & 0.4814$\pm$0.0098 & 0.4648$\pm$0.0083 & 0.4771$\pm$0.0081 & 0.4734$\pm$0.0089 & 0.4769$\pm$0.0090 & 0.4932$\pm$0.0090 \\
{[\ion{O}{3}]$\lambda$4363  } & 0.0467$\pm$0.0048 & 0.0606$\pm$0.0025 & 0.0818$\pm$0.0015 & 0.1288$\pm$0.0023 & 0.1735$\pm$0.0029 & 0.1276$\pm$0.0022 & 0.1827$\pm$0.0032 \\
{\ion{He}{1} $\lambda$4388  } & \nodata & 0.0026$\pm$0.0023 & 0.0042$\pm$0.0007 & 0.0045$\pm$0.0010 & 0.0044$\pm$0.0003 & 0.0045$\pm$0.0004 & 0.0043$\pm$0.0014 \\
{\ion{He}{1} $\lambda$4438  } & 0.0034$\pm$0.0042 & 0.0011$\pm$0.0027 & 0.0002$\pm$0.0006 & 0.0008$\pm$0.0010 & 0.0004$\pm$0.0003 & 0.0009$\pm$0.0004 & \nodata \\
{\ion{He}{1} $\lambda$4471  } & 0.0313$\pm$0.0042 & 0.0331$\pm$0.0028 & 0.0363$\pm$0.0008 & 0.0374$\pm$0.0011 & 0.0381$\pm$0.0007 & 0.0384$\pm$0.0007 & 0.0429$\pm$0.0016 \\
{\ion{Mg}{1}] $\lambda$4563  } & \nodata & 0.0025$\pm$0.0026 & 0.0021$\pm$0.0007 & 0.0011$\pm$0.0007 & 0.0013$\pm$0.0003 & 0.0017$\pm$0.0003 & 0.0010$\pm$0.0008 \\
{\ion{Mg}{1}] $\lambda$4571  } & 0.0006$\pm$0.0041 & 0.0014$\pm$0.0026 & 0.0011$\pm$0.0007 & 0.0014$\pm$0.0007 & 0.0009$\pm$0.0003 & 0.0015$\pm$0.0003 & 0.0008$\pm$0.0008 \\
{\ion{O}{2} $\lambda$4639+4642  } & 0.0035$\pm$0.0025 & 0.0039$\pm$0.0019 & 0.0020$\pm$0.0007 & \nodata & 0.0005$\pm$0.0003 & 0.0010$\pm$0.0004 & 0.0027$\pm$0.0009 \\
{\ion{O}{2} $\lambda$4649+4651  } & 0.0034$\pm$0.0025 & 0.0008$\pm$0.0019 & 0.0015$\pm$0.0007 & 0.0006$\pm$0.0006 & 0.0006$\pm$0.0003 & 0.0011$\pm$0.0004 & 0.0013$\pm$0.0009 \\
{[\ion{Fe}{3}] $\lambda$4658  } & 0.0007$\pm$0.0025 & 0.0038$\pm$0.0019 & 0.0050$\pm$0.0007 & 0.0034$\pm$0.0006 & 0.0022$\pm$0.0003 & 0.0017$\pm$0.0004 & 0.0080$\pm$0.0009 \\
{\ion{O}{2} $\lambda$4662  } & \nodata & 0.0003$\pm$0.0019 & 0.0008$\pm$0.0007 & 0.0005$\pm$0.0006 & 0.0006$\pm$0.0003 & 0.0002$\pm$0.0004 & 0.0015$\pm$0.0009 \\
{\ion{He}{2} $\lambda$4686  } & \nodata & 0.0276$\pm$0.0019 & 0.0087$\pm$0.0005 & 0.0100$\pm$0.0006 & 0.0059$\pm$0.0003 & 0.0019$\pm$0.0004 & 0.0198$\pm$0.0010 \\
{[\ion{Fe}{3}] $\lambda$4702  } & 0.0030$\pm$0.0025 & 0.0002$\pm$0.0019 & 0.0032$\pm$0.0007 & 0.0010$\pm$0.0006 & 0.0013$\pm$0.0003 & 0.0008$\pm$0.0004 & 0.0020$\pm$0.0009 \\
{[\ion{Ar}{4}] $\lambda$4711  } & 0.0014$\pm$0.0025 & 0.0030$\pm$0.0019 & 0.0099$\pm$0.0007 & 0.0133$\pm$0.0006 & 0.0333$\pm$0.0006 & 0.0149$\pm$0.0004 & 0.0157$\pm$0.0009 \\
{\ion{He}{1} $\lambda$4713  } & 0.0045$\pm$0.0025 & 0.0041$\pm$0.0019 & 0.0062$\pm$0.0007 & 0.0047$\pm$0.0006 & 0.0043$\pm$0.0003 & 0.0052$\pm$0.0004 & 0.0073$\pm$0.0009 \\
{[\ion{Ar}{4}] $\lambda$4740  } & 0.0039$\pm$0.0025 & 0.0019$\pm$0.0019 & 0.0078$\pm$0.0007 & 0.0086$\pm$0.0006 & 0.0253$\pm$0.0005 & 0.0120$\pm$0.0004 & 0.0141$\pm$0.0009 \\
{\ion{Fe}{2} $\lambda$4827  } & \nodata & \nodata & 0.0010$\pm$0.0014 & \nodata & \nodata & \nodata & \nodata \\
{H$\beta$ $\lambda$4861  } & 1.0000$\pm$0.0173 & 1.0000$\pm$0.0168 & 1.0000$\pm$0.0155 & 1.0000$\pm$0.0151 & 1.0000$\pm$0.0163 & 1.0000$\pm$0.0162 & 1.0000$\pm$0.0156 \\
{\ion{Fe}{2} $\lambda$4883  } & \nodata & \nodata & 0.0008$\pm$0.0014 & 0.0014$\pm$0.0009 & 0.0005$\pm$0.0003 & 0.0011$\pm$0.0006 & 0.0031$\pm$0.0018 \\
{\ion{Si}{2} $\lambda$4903  } & \nodata & \nodata & 0.0003$\pm$0.0006 & 0.0004$\pm$0.0006 & 0.0003$\pm$0.0004 & 0.0002$\pm$0.0002 & 0.0013$\pm$0.0019 \\
{[\ion{Fe}{2}] $\lambda$4905  } & 0.0005$\pm$0.0036 & \nodata & 0.0001$\pm$0.0006 & 0.0004$\pm$0.0006 & 0.0007$\pm$0.0004 & 0.0004$\pm$0.0002 & 0.0030$\pm$0.0019 \\
{\ion{He}{1} $\lambda$4922  } & 0.0058$\pm$0.0036 & 0.0082$\pm$0.0020 & 0.0101$\pm$0.0006 & 0.0102$\pm$0.0006 & 0.0098$\pm$0.0004 & 0.0094$\pm$0.0003 & 0.0120$\pm$0.0019 \\
{[\ion{O}{3}] $\lambda$4959  } & 0.4732$\pm$0.0085 & 0.6691$\pm$0.0111 & 2.0008$\pm$0.0305 & 1.5439$\pm$0.0232 & 2.5590$\pm$0.0411 & 2.2432$\pm$0.0358 & 1.8672$\pm$0.0288 \\
{\ion{Fe}{2} $\lambda$4977  } & 0.0018$\pm$0.0036 & 0.0005$\pm$0.0020 & 0.0005$\pm$0.0006 & \nodata & 0.0004$\pm$0.0004 & 0.0001$\pm$0.0002 & 0.0006$\pm$0.0018 \\
{[\ion{Fe}{3}] $\lambda$4986  } & 0.0047$\pm$0.0036 & 0.0062$\pm$0.0021 & 0.0044$\pm$0.0006 & 0.0042$\pm$0.0006 & 0.0031$\pm$0.0004 & 0.0007$\pm$0.0002 & 0.0037$\pm$0.0018 \\
{\ion{Fe}{2} $\lambda$4991  } & 0.0002$\pm$0.0036 & \nodata & \nodata & \nodata & \nodata & \nodata & \nodata \\
{[\ion{O}{3}] $\lambda$5007  } & 1.3760$\pm$0.0227 & 1.9845$\pm$0.0325 & 6.0061$\pm$0.0918 & 4.5676$\pm$0.0690 & 7.3659$\pm$0.1188 & 6.9190$\pm$0.1108 & 5.5926$\pm$0.0865 \\
{\ion{He}{1} $\lambda$5016  } & 0.0209$\pm$0.0036 & 0.0190$\pm$0.0021 & 0.0195$\pm$0.0007 & 0.0198$\pm$0.0006 & 0.0169$\pm$0.0005 & 0.0185$\pm$0.0004 & 0.0246$\pm$0.0019 \\
{\ion{Si}{2} $\lambda$5041  } & 0.0018$\pm$0.0036 & 0.0020$\pm$0.0021 & 0.0014$\pm$0.0006 & 0.0025$\pm$0.0006 & 0.0010$\pm$0.0004 & 0.0006$\pm$0.0002 & 0.0026$\pm$0.0018 \\
{\ion{He}{1} $\lambda$5048  } & 0.0034$\pm$0.0036 & 0.0010$\pm$0.0021 & 0.0018$\pm$0.0006 & 0.0025$\pm$0.0006 & 0.0018$\pm$0.0004 & 0.0016$\pm$0.0002 & 0.0021$\pm$0.0018 \\
{\ion{Si}{2} $\lambda$5056  } & \nodata & 0.0022$\pm$0.0021 & 0.0013$\pm$0.0006 & 0.0010$\pm$0.0006 & 0.0004$\pm$0.0004 & 0.0003$\pm$0.0002 & 0.0020$\pm$0.0018 \\
{[\ion{Ar}{3}] $\lambda$5192  } & \nodata & \nodata & 0.0009$\pm$0.0005 & \nodata & 0.0009$\pm$0.0003 & 0.0008$\pm$0.0003 & 0.0003$\pm$0.0011 \\
{[\ion{N}{1}] $\lambda$5198  } & 0.0029$\pm$0.0027 & 0.0026$\pm$0.0028 & 0.0016$\pm$0.0005 & 0.0040$\pm$0.0007 & 0.0009$\pm$0.0003 & 0.0022$\pm$0.0003 & 0.0032$\pm$0.0011 \\
{[\ion{N}{1}] $\lambda$5200  } & \nodata & 0.0000$\pm$0.0028 & 0.0004$\pm$0.0005 & 0.0007$\pm$0.0007 & 0.0010$\pm$0.0003 & 0.0006$\pm$0.0003 & \nodata \\
{[\ion{Fe}{3}] $\lambda$5270  } & \nodata & 0.0005$\pm$0.0034 & 0.0016$\pm$0.0004 & 0.0016$\pm$0.0006 & 0.0011$\pm$0.0003 & 0.0006$\pm$0.0003 & 0.0042$\pm$0.0010 \\
{[\ion{Cl}{3}] $\lambda$5518  } & \nodata & 0.0048$\pm$0.0047 & 0.0038$\pm$0.0008 & 0.0020$\pm$0.0011 & 0.0024$\pm$0.0002 & 0.0035$\pm$0.0003 & 0.0017$\pm$0.0003 \\
{[\ion{Cl}{3}] $\lambda$5538  } & 0.0018$\pm$0.0226 & 0.0032$\pm$0.0047 & \nodata & 0.0022$\pm$0.0011 & 0.0015$\pm$0.0002 & 0.0021$\pm$0.0003 & 0.0016$\pm$0.0003 \\
{[\ion{N}{2}] $\lambda$5755  } & 0.0015$\pm$0.0025 & 0.0004$\pm$0.0010 & 0.0016$\pm$0.0012 & 0.0005$\pm$0.0003 & 0.0006$\pm$0.0002 & 0.0012$\pm$0.0001 & 0.0039$\pm$0.0006 \\
{\ion{He}{1} $\lambda$5876  } & 0.1018$\pm$0.0040 & 0.0914$\pm$0.0022 & 0.1047$\pm$0.0022 & 0.1034$\pm$0.0019 & 0.1016$\pm$0.0018 & 0.1115$\pm$0.0020 & 0.1194$\pm$0.0021 \\
{\ion{Si}{2} $\lambda$5958  } & \nodata & \nodata & 0.0002$\pm$0.0003 & 0.0007$\pm$0.0003 & 0.0004$\pm$0.0001 & 0.0005$\pm$0.0001 & 0.0007$\pm$0.0004 \\
{\ion{Si}{2} $\lambda$5978  } & 0.0005$\pm$0.0037 & 0.0018$\pm$0.0012 & 0.0003$\pm$0.0003 & 0.0005$\pm$0.0003 & 0.0002$\pm$0.0001 & 0.0004$\pm$0.0001 & 0.0008$\pm$0.0004 \\
{[\ion{K}{4}] $\lambda$6101  } & 0.0029$\pm$0.0032 & \nodata & 0.0004$\pm$0.0003 & 0.0006$\pm$0.0003 & 0.0012$\pm$0.0001 & 0.0006$\pm$0.0001 & 0.0006$\pm$0.0004 \\
{[\ion{O}{1}] $\lambda$6300  } & 0.0025$\pm$0.0036 & 0.0073$\pm$0.0009 & 0.0140$\pm$0.0004 & 0.0190$\pm$0.0005 & 0.0146$\pm$0.0003 & 0.0196$\pm$0.0004 & 0.0167$\pm$0.0005 \\
{[\ion{S}{3}] $\lambda$6312  } & 0.0061$\pm$0.0036 & 0.0070$\pm$0.0009 & 0.0215$\pm$0.0005 & 0.0109$\pm$0.0004 & 0.0127$\pm$0.0003 & 0.0181$\pm$0.0004 & 0.0112$\pm$0.0005 \\
{\ion{Si}{2} $\lambda$6347  } & \nodata & 0.0014$\pm$0.0001 & 0.0001$\pm$0.0003 & 0.0001$\pm$0.0003 & 0.0003$\pm$0.0001 & \nodata & 0.0007$\pm$0.0002 \\
{[\ion{O}{1}] $\lambda$6364  } & 0.0039$\pm$0.0028 & 0.0024$\pm$0.0009 & 0.0043$\pm$0.0003 & 0.0063$\pm$0.0003 & 0.0047$\pm$0.0002 & 0.0064$\pm$0.0002 & 0.0052$\pm$0.0006 \\
{\ion{Si}{2} $\lambda$6371  } & 0.0008$\pm$0.0028 & 0.0006$\pm$0.0009 & 0.0005$\pm$0.0003 & 0.0012$\pm$0.0003 & 0.0005$\pm$0.0001 & 0.0006$\pm$0.0001 & 0.0010$\pm$0.0006 \\
{[\ion{N}{2}] $\lambda$6548  } & 0.0021$\pm$0.0021 & 0.0029$\pm$0.0011 & 0.0151$\pm$0.0004 & 0.0052$\pm$0.0002 & 0.0072$\pm$0.0002 & 0.0149$\pm$0.0003 & 0.0231$\pm$0.0005 \\
{H$\alpha$ $\lambda$6563  } & 2.7585$\pm$0.0570 & 2.7951$\pm$0.0569 & 2.8037$\pm$0.0515 & 2.7674$\pm$0.0510 & 2.7766$\pm$0.0543 & 2.8934$\pm$0.0567 & 2.9489$\pm$0.0562 \\
{[\ion{N}{2}] $\lambda$6584  } & 0.0177$\pm$0.0021 & 0.0108$\pm$0.0011 & 0.0462$\pm$0.0009 & 0.0179$\pm$0.0004 & 0.0229$\pm$0.0005 & 0.0477$\pm$0.0009 & 0.0761$\pm$0.0015 \\
{\ion{He}{1} $\lambda$6678  } & 0.0306$\pm$0.0020 & 0.0256$\pm$0.0008 & 0.0299$\pm$0.0006 & 0.0291$\pm$0.0006 & 0.0285$\pm$0.0006 & 0.0315$\pm$0.0006 & 0.0290$\pm$0.0006 \\
{[\ion{S}{2}] $\lambda$36717  } & 0.0310$\pm$0.0029 & 0.0314$\pm$0.0010 & 0.0924$\pm$0.0017 & 0.0606$\pm$0.0012 & 0.0568$\pm$0.0013 & 0.0817$\pm$0.0016 & 0.0357$\pm$0.0008 \\
{[\ion{S}{2}] $\lambda$6731  } & 0.0224$\pm$0.0029 & 0.0230$\pm$0.0009 & 0.0676$\pm$0.0013 & 0.0459$\pm$0.0009 & 0.0426$\pm$0.0011 & 0.0616$\pm$0.0012 & 0.0326$\pm$0.0007 \\
{[\ion{Fe}{4}] $\lambda$6740  } & \nodata & 0.0015$\pm$0.0008 & 0.0002$\pm$0.0002 & 0.0006$\pm$0.0003 & \nodata & 0.0003$\pm$0.0001 & 0.0014$\pm$0.0003 \\
{\ion{Ni}{2} $\lambda$6794  } & \nodata & \nodata & \nodata & 0.0001$\pm$0.0003 & 0.0002$\pm$0.0009 & 0.0001$\pm$0.0001 & 0.0004$\pm$0.0004 \\
{\ion{He}{1} $\lambda$7065  } & 0.0224$\pm$0.0040 & 0.0221$\pm$0.0012 & 0.0242$\pm$0.0005 & 0.0341$\pm$0.0008 & 0.0284$\pm$0.0006 & 0.0311$\pm$0.0007 & 0.0641$\pm$0.0014 \\
{[\ion{Ar}{3}] $\lambda$7136  } & 0.0141$\pm$0.0045 & 0.0188$\pm$0.0011 & 0.0728$\pm$0.0015 & 0.0319$\pm$0.0008 & 0.0511$\pm$0.0011 & 0.0746$\pm$0.0019 & 0.0325$\pm$0.0009 \\
{[\ion{Fe}{3}] $\lambda$7155  } & \nodata & 0.0001$\pm$0.0010 & \nodata & 0.0005$\pm$0.0005 & 0.0001$\pm$0.0001 & 0.0001$\pm$0.0001 & 0.0008$\pm$0.0006 \\
{[\ion{Mg}{4}] $\lambda$7170  } & \nodata & \nodata & 0.0003$\pm$0.0004 & \nodata & 0.0014$\pm$0.0001 & 0.0005$\pm$0.0001 & 0.0010$\pm$0.0006 \\
{\ion{He}{1} $\lambda$7281  } & 0.0031$\pm$0.0040 & 0.0051$\pm$0.0011 & 0.0051$\pm$0.0005 & 0.0066$\pm$0.0005 & 0.0059$\pm$0.0002 & 0.0063$\pm$0.0002 & 0.0053$\pm$0.0048 \\
{\ion{Ca}{2} $\lambda$7291  } & 0.0008$\pm$0.0040 & 0.0004$\pm$0.0011 & \nodata & 0.0005$\pm$0.0005 & 0.0000$\pm$0.0001 & 0.0000$\pm$0.0002 & \nodata \\
{[\ion{O}{2}] $\lambda$7320  } & 0.0072$\pm$0.0042 & 0.0060$\pm$0.0016 & 0.0136$\pm$0.0004 & 0.0123$\pm$0.0004 & 0.0075$\pm$0.0002 & 0.0154$\pm$0.0004 & \nodata \\
{[\ion{O}{2}] $\lambda$7330  } & 0.0060$\pm$0.0042 & 0.0036$\pm$0.0016 & 0.0111$\pm$0.0003 & 0.0100$\pm$0.0004 & 0.0065$\pm$0.0002 & 0.0130$\pm$0.0004 & 0.0009$\pm$0.0045 \\
{[\ion{Ni}{2}] $\lambda$7378  } & 0.0012$\pm$0.0041 & 0.0017$\pm$0.0018 & 0.0004$\pm$0.0003 & 0.0009$\pm$0.0002 & 0.0003$\pm$0.0001 & 0.0003$\pm$0.0002 & 0.0002$\pm$0.0020 \\
{[\ion{Cl}{4}] $\lambda$7531  } & 0.0005$\pm$0.0023 & 0.0005$\pm$0.0009 & 0.0007$\pm$0.0068 & \nodata & 0.0010$\pm$0.0008 & 0.0010$\pm$0.0001 & 0.0004$\pm$0.0003 \\
{[\ion{Ar}{3}] $\lambda$7751  } & 0.0033$\pm$0.0027 & 0.0043$\pm$0.0022 & 0.0232$\pm$0.0006 & 0.0072$\pm$0.0004 & 0.0122$\pm$0.0003 & 0.0171$\pm$0.0004 & 0.0071$\pm$0.0003 \\
{[\ion{Ar}{3}] $\lambda$8037  } & \nodata & 0.0013$\pm$0.0009 & 0.0001$\pm$0.0002 & \nodata & 0.0002$\pm$0.0002 & \nodata & \nodata \\
{[\ion{Cl}{4}] $\lambda$8046  } & 0.0006$\pm$0.0027 & 0.0017$\pm$0.0009 & 0.0011$\pm$0.0002 & 0.0017$\pm$0.0003 & 0.0033$\pm$0.0002 & 0.0024$\pm$0.0001 & 0.0009$\pm$0.0003 \\
{P19 $\lambda$8413} & \nodata & 0.0039$\pm$0.0019 & 0.0032$\pm$0.0002 & 0.0023$\pm$0.0002 & 0.0028$\pm$0.0002 & 0.0033$\pm$0.0004 & 0.0027$\pm$0.0006 \\
{P18 $\lambda$8438} & 0.0001$\pm$0.0046 & 0.0021$\pm$0.0019 & 0.0029$\pm$0.0002 & 0.0029$\pm$0.0002 & 0.0032$\pm$0.0003 & 0.0032$\pm$0.0004 & 0.0036$\pm$0.0006 \\
{\ion{O}{1} $\lambda$8446} & \nodata & 0.0051$\pm$0.0019 & 0.0022$\pm$0.0002 & 0.0055$\pm$0.0003 & 0.0018$\pm$0.0002 & 0.0026$\pm$0.0004 & 0.0201$\pm$0.0009 \\
{P17 $\lambda$8467} & \nodata & 0.0037$\pm$0.0019 & 0.0040$\pm$0.0002 & 0.0029$\pm$0.0002 & 0.0034$\pm$0.0003 & 0.0038$\pm$0.0004 & 0.0047$\pm$0.0006 \\
{P16 $\lambda$8502} & \nodata & 0.0042$\pm$0.0015 & 0.0046$\pm$0.0003 & 0.0041$\pm$0.0002 & 0.0042$\pm$0.0003 & 0.0046$\pm$0.0003 & 0.0042$\pm$0.0007 \\
{P15 $\lambda$8545} & \nodata & 0.0052$\pm$0.0009 & 0.0041$\pm$0.0005 & 0.0038$\pm$0.0005 & 0.0051$\pm$0.0002 & 0.0057$\pm$0.0002 & 0.0044$\pm$0.0010 \\
{\ion{Cl}{2} $\lambda$8579} & \nodata & 0.0007$\pm$0.0010 & 0.0003$\pm$0.0006 & 0.0008$\pm$0.0005 & 0.0003$\pm$0.0001 & 0.0006$\pm$0.0002 & 0.0003$\pm$0.0008 \\
{\ion{He}{1} $\lambda$8583} & \nodata & \nodata & 0.0007$\pm$0.0006 & 0.0003$\pm$0.0005 & 0.0005$\pm$0.0001 & 0.0006$\pm$0.0002 & \nodata \\
{P14 $\lambda$8598} & \nodata & 0.0073$\pm$0.0010 & 0.0058$\pm$0.0006 & 0.0056$\pm$0.0005 & 0.0066$\pm$0.0002 & 0.0068$\pm$0.0003 & 0.0067$\pm$0.0008 \\
{\ion{He}{1} $\lambda$8617} & 0.0195$\pm$0.0110 & \nodata & \nodata & 0.0006$\pm$0.0005 & 0.0002$\pm$0.0001 & 0.0000$\pm$0.0002 & 0.0006$\pm$0.0008 \\
{\ion{He}{1} $\lambda$8635} & 0.0090$\pm$0.0091 & 0.0006$\pm$0.0011 & 0.0002$\pm$0.0007 & \nodata & \nodata & \nodata & 0.0008$\pm$0.0004 \\
{P13 $\lambda$8665} & 0.0220$\pm$0.0091 & 0.0085$\pm$0.0011 & 0.0073$\pm$0.0007 & 0.0076$\pm$0.0004 & 0.0078$\pm$0.0003 & 0.0085$\pm$0.0003 & 0.0075$\pm$0.0005 \\
{\ion{N}{1} $\lambda$8683} & \nodata & 0.0010$\pm$0.0011 & 0.0005$\pm$0.0007 & \nodata & 0.0005$\pm$0.0002 & 0.0006$\pm$0.0002 & 0.0007$\pm$0.0004 \\
{\ion{Fe}{2} $\lambda$8733} & \nodata & 0.0003$\pm$0.0040 & 0.0006$\pm$0.0007 & 0.0005$\pm$0.0006 & 0.0006$\pm$0.0002 & 0.0004$\pm$0.0015 & \nodata \\
{P12 $\lambda$8750} & \nodata & 0.0086$\pm$0.0040 & 0.0102$\pm$0.0007 & 0.0095$\pm$0.0006 & 0.0105$\pm$0.0004 & 0.0111$\pm$0.0015 & 0.0088$\pm$0.0004 \\
{[\ion{S}{3}] $\lambda$8829} & 0.0038$\pm$0.0085 & 0.0008$\pm$0.0051 & \nodata & 0.0000$\pm$0.0006 & 0.0002$\pm$0.0002 & 0.0001$\pm$0.0019 & \nodata \\
{P11 $\lambda$8863} & 0.0129$\pm$0.0085 & 0.0136$\pm$0.0051 & 0.0126$\pm$0.0006 & 0.0121$\pm$0.0007 & 0.0134$\pm$0.0005 & 0.0139$\pm$0.0019 & 0.0122$\pm$0.0005 \\
{P10 $\lambda$9015} & 0.0189$\pm$0.0032 & 0.0178$\pm$0.0018 & 0.0159$\pm$0.0005 & 0.0186$\pm$0.0006 & 0.0173$\pm$0.0006 & 0.0177$\pm$0.0006 & 0.0149$\pm$0.0008 \\
{[\ion{S}{3}] $\lambda$9069} & 0.0281$\pm$0.0045 & 0.0394$\pm$0.0021 & 0.2030$\pm$0.0055 & 0.0658$\pm$0.0018 & 0.0817$\pm$0.0030 & 0.1393$\pm$0.0048 & 0.0687$\pm$0.0027 \\
{[\ion{Cl}{2}] $\lambda$9124} & 0.0017$\pm$0.0042 & \nodata & 0.0004$\pm$0.0005 & \nodata & 0.0002$\pm$0.0004 & \nodata & 0.0004$\pm$0.0013 \\
{\ion{He}{1} $\lambda$9210} & 0.0018$\pm$0.0033 & 0.0001$\pm$0.0011 & 0.0007$\pm$0.0016 & 0.0013$\pm$0.0026 & 0.0002$\pm$0.0007 & 0.0010$\pm$0.0002 & 0.0010$\pm$0.0013 \\
{P9 $\lambda$9229} & 0.0174$\pm$0.0034 & 0.0252$\pm$0.0015 & 0.0213$\pm$0.0017 & 0.0178$\pm$0.0027 & 0.0242$\pm$0.0012 & 0.0247$\pm$0.0009 & 0.0220$\pm$0.0016 \\
{[\ion{S}{3}] $\lambda$9531} & 0.0609$\pm$0.0151 & 0.1140$\pm$0.0072 & 0.4690$\pm$0.0141 & 0.1621$\pm$0.0051 & 0.1850$\pm$0.0096 & 0.3652$\pm$0.0164 & 0.1416$\pm$0.0076 \\
{P8 $\lambda$9546} & 0.0345$\pm$0.0150 & 0.0325$\pm$0.0051 & 0.0387$\pm$0.0016 & 0.0353$\pm$0.0014 & 0.0327$\pm$0.0018 & 0.0357$\pm$0.0017 & 0.0267$\pm$0.0016 \\
\enddata
\tablecomments{Emission lines measured in the MODS spectra. Columns provide all emission line intensities in the MODS data for a single target. Each row includes the reddening corrected line intensity relative to H$\beta$ and uncertainty in the ratio, considering the uncertainty on both the line flux and $E(B-V)$. Spectroscopic redshift, $E(B-V)$ used to reddening correct the spectrum, and $a_H$ can be found in Table \ref{t:hiinten}. Note that significantly detected lines can be rejected for \te, \den, and abundance analysis if coincident with strong telluric absorption features (see \S\ref{s:phys}). Table truncated at seven targets, the complete version is available in machine readable format.}
\end{deluxetable*}


\startlongtable
\begin{deluxetable*}{lcccccccc}  
\tablecaption{LBT \yp\ Project Physical Conditions and Oxygen Abundances \label{t:abun}}
\tablewidth{\columnwidth}
\tabletypesize{\footnotesize}
\tablehead{
   \colhead{Property} & 
   \colhead{AGC198691} & 
   \colhead{DDO68} & 
   \colhead{HS0029+1748} &
   \colhead{HS0122+0743} &
   \colhead{HS0134+3415} &
   \colhead{HS0811+4913} &
   \colhead{HS0837+4717}}
\startdata
n$_e$[\ion{S}{2}] (cm$^{-3}$)   &  35$_{- 20}^{+570}$ &  55$_{- 29}^{+ 62}$ &  56$_{- 12}^{+ 13}$ &  95$_{- 16}^{+ 18}$ &  83$_{- 25}^{+ 27}$ &  91$_{- 13}^{+ 13}$ & 431$_{- 36}^{+ 38}$ \\
n$_e$[\ion{O}{2}] (cm$^{-3}$)   &  21$_{- 12}^{+ 94}$ &   9$_{-  5}^{+ 50}$ &  33$_{- 12}^{+ 13}$ &  55$_{- 15}^{+ 15}$ & 104$_{- 15}^{+ 16}$ & 151$_{- 16}^{+ 18}$ & 546$_{- 41}^{+ 46}$ \\
n$_e$[\ion{Cl}{3}] (cm$^{-3}$)   & \nodata & \nodata & \nodata & \nodata & \nodata & \nodata & 2205$_{-1270}^{+12760}$ \\
n$_e$[\ion{Ar}{4}] (cm$^{-3}$)   & \nodata & \nodata & 680$_{-391}^{+1176}$ & \nodata & 172$_{- 91}^{+160}$ & 851$_{-348}^{+407}$ & 1913$_{-840}^{+1298}$ \\
\vspace{-0.30cm} \\
n$_{e,Used}$ (cm$^{-3}$)   & 100$_{- 50}^{+ 50}$ & 100$_{- 50}^{+ 50}$ & 100$_{- 50}^{+ 50}$ & 100$_{- 50}^{+ 50}$ & 100$_{- 50}^{+ 50}$ & 100$_{- 50}^{+ 50}$ & 488$_{- 27}^{+ 30}$ \\
\vspace{-0.15cm} \\
T$_e$[\ion{N}{2}] (K)   & \nodata & \nodata & \nodata & \nodata & 13200$\pm$1700 & 12600$\pm$ 700 & 19800$\pm$2400 \\
T$_e$[\ion{O}{2}] (K)   & \nodata & 13000$\pm$3400 & 15400$\pm$ 400 & 15600$\pm$ 500 & 16200$\pm$ 500 & 12900$\pm$ 300 & \nodata \\
T$_e$[\ion{S}{2}] (K)   & \nodata & \nodata & \nodata & 19200$\pm$2700 & 15800$\pm$1200 & 15400$\pm$1100 & \nodata \\
T$_e$[\ion{S}{3}] (K)   & \nodata & 17000$\pm$1600 & 12400$\pm$ 200 & 16100$\pm$ 400 & 15500$\pm$ 300 & 13900$\pm$ 200 & 18400$\pm$ 900 \\
T$_e$[\ion{Ar}{3}] (K)   & \nodata & \nodata & \nodata & \nodata & 14900$\pm$2600 & \nodata & \nodata \\
T$_e$[\ion{O}{3}] (K)   & 20000$\pm$1300 & 18900$\pm$ 500 & 12900$\pm$ 100 & 17800$\pm$ 200 & 16300$\pm$ 200 & 14600$\pm$ 200 & 19500$\pm$ 300 \\
\vspace{-0.40cm} \\
T$_{e,Low}$ (K)   & 17000$\pm$1100 & 16200$\pm$ 700 & 12000$\pm$ 600 & 15500$\pm$ 600 & 13200$\pm$1700 & 12600$\pm$ 700 & 19800$\pm$2400 \\
T$_{e,Int}$ (K)   & 18500$\pm$1200 & 17000$\pm$1600 & 12400$\pm$ 200 & 16100$\pm$ 400 & 15500$\pm$ 300 & 13900$\pm$ 200 & 18400$\pm$ 900 \\
T$_{e,High}$ (K)   & 20000$\pm$1300 & 18900$\pm$ 500 & 12900$\pm$ 100 & 17800$\pm$ 200 & 16300$\pm$ 200 & 14600$\pm$ 200 & 19500$\pm$ 300 \\
\vspace{-0.15cm} \\
O$^{+}$/H$^+$ ($\times$10$^5$)   & 0.24$\pm$0.05 & 0.27$\pm$0.03 & 1.43$\pm$0.28 & 0.53$\pm$0.07 & 0.51$\pm$0.25 & 1.54$\pm$0.33 & 0.21$\pm$0.07 \\
O$^{2+}$/H$^+$ ($\times$10$^5$)   & 0.77$\pm$0.11 & 1.25$\pm$0.07 & 9.74$\pm$0.32 & 3.27$\pm$0.11 & 6.42$\pm$0.19 & 7.98$\pm$0.25 & 3.26$\pm$0.12 \\
12+log(O/H)   & 7.002$\pm$0.050 & 7.181$\pm$0.022 & 8.048$\pm$0.016 & 7.580$\pm$0.015 & 7.841$\pm$0.020 & 7.979$\pm$0.019 & 7.541$\pm$0.017 \\
\enddata
\tablecomments{Physical conditions, ionic and total O abundances measured in the \yp\ Project nebulae. Each row corresponds to a property measured in a given target (columns). These properties are: Direct \den\ measurements (Rows 1-4), adopted \den\ in the ISM (5, described in \S\ref{s:dentrends}), direct \te\ (6-11), adopted \te\ for each ionization zone (12-14), ionic abundances of O$^+$ and O$^{2+}$ (15-16), and the total O abundance, 12+log(O/H) (17). Table truncated at seven targets, the complete version is available in machine readable format.}
\end{deluxetable*}

\bibliographystyle{aasjournalv7}
\bibliography{yp_total_bib}

\clearpage

\end{document}